\documentclass[12pt,dvips]{article}
\usepackage{rotating}
\usepackage{amsmath}
\usepackage{mathtools}
\usepackage{amsthm}
\usepackage{amsfonts}
\usepackage[dvips]{epsfig}
\usepackage{graphicx}
\usepackage{amssymb}
\usepackage{cancel}
\usepackage{pstricks} 
\usepackage{lscape}
\usepackage{cancel}
\usepackage{slashed}
\usepackage{pstricks} 
\usepackage{lscape}
\usepackage{color}
\usepackage{cite}
\newcommand{\beq}{\begin{equation}}
\newcommand{\eeq}{\end{equation}}
\newcommand{\ga}{\lower.7ex\hbox{$\;\stackrel{\textstyle>}{\sim}\;$}}
\newcommand{\la}{\lower.7ex\hbox{$\;\stackrel{\textstyle<}{\sim}\;$}}

\newcommand{\vphi}{\varphi}

\newcommand{\kahler}{K\"ahler }

\begin{document}

\def\thefootnote{\fnsymbol{footnote}}

\begin{flushright}
{\tt KCL-PH-TH/2015-14}, {\tt LCTS/2015-06}, {\tt CERN-PH-TH/2015-064}  \\
{\tt ACT-03-15, UMN-TH-3426/15, FTPI-MINN-15/12} \\
\end{flushright}

\vspace{1cm}
\begin{center}
{\bf {\large Phenomenological Aspects of No-Scale Inflation Models}}
\vspace {0.1in}
\end{center}

\vspace{0.05in}

\begin{center}{
{\bf John~Ellis}$^{a}$,
{\bf Marcos~A.~G.~Garcia}$^{b}$,
{\bf Dimitri~V.~Nanopoulos}$^{c}$ and
{\bf Keith~A.~Olive}$^{b}$
}
\end{center}

\begin{center}
{\em $^a$Theoretical Particle Physics and Cosmology Group, Department of
  Physics, King's~College~London, London WC2R 2LS, United Kingdom;\\
Theory Division, CERN, CH-1211 Geneva 23,
  Switzerland}\\[0.2cm]
  {\em $^b$William I. Fine Theoretical Physics Institute, School of Physics and Astronomy,\\
University of Minnesota, Minneapolis, MN 55455, USA}\\[0.2cm]
{\em $^c$George P. and Cynthia W. Mitchell Institute for Fundamental Physics and Astronomy,
Texas A\&M University, College Station, TX 77843, USA;\\
Astroparticle Physics Group, Houston Advanced Research Center (HARC), \\ Mitchell Campus, Woodlands, TX 77381, USA;\\
Academy of Athens, Division of Natural Sciences,
Athens 10679, Greece}\\
\end{center}

\bigskip

\centerline{\bf ABSTRACT}

\noindent  
{\small We discuss phenomenological aspects of no-scale supergravity inflationary models motivated by compactified string
models, in which the inflaton may be identified either as a K\"ahler modulus or an untwisted matter field,
focusing on models that make predictions for the scalar spectral index $n_s$ and the
tensor-to-scalar ratio $r$ that are similar to the Starobinsky model. We discuss possible patterns of soft supersymmetry breaking,
exhibiting examples of the pure no-scale type $m_0 = B_0 = A_0 = 0$, of the CMSSM type
with universal $A_0$ and $m_0 \ne 0$ at a high scale, and of the mSUGRA type with $A_0 = B_0 + m_0$ 
boundary conditions at the high input scale. These may be combined with
a non-trivial gauge kinetic function that generates gaugino masses $m_{1/2} \ne 0$,
or one may have a pure gravity mediation scenario where trilinear terms and gaugino masses are generated through anomalies.
We also discuss inflaton decays and reheating, showing possible decay channels for the inflaton when it is 
either an
untwisted matter field or a K\"ahler modulus. Reheating is very efficient if a matter field inflaton is directly coupled to MSSM fields, and both candidates lead to sufficient reheating in the presence of a non-trivial gauge kinetic function.}

\vspace{0.2in}

\begin{flushleft}
March 2015
\end{flushleft}
\medskip
\noindent

\newpage

\section{Introduction}

One of the biggest challenges for string theory is how to connect with particle and/or
cosmological experiments. No-scale supergravity \cite{no-scale,LN} is the natural framework to seek such
connections, for many reasons. On the one hand, physics at scales hierarchically smaller
than the Planck scale is expected to be protected by approximate supersymmetry,
which should be local and combined with gravity in some supergravity theory. No-scale
supergravity is the most appropriate form, since it emerges naturally as the
effective low-energy theory derived from compactified string \cite{Witten1985}, and yields a positive
semi-definite potential at the tree level, thus lending itself naturally to cosmology.
Moreover, no-scale supergravity has recently emerged as a very effective
framework for models of cosmological inflation, as discussed in~\cite{EENOS,enq,gl2,bg,msy2,Davis:2008fv,Ant,ADEH,ENO6,ENO7,FKR,buch,ENO8,Pallis,FeKR,EGNO,EGNO2,EGNO3,Li,KT,bdhw,klt,LT}, 
yielding models whose predictions
can interpolate between those of the Starobinsky model \cite{Staro} and chaotic inflation with a
quadratic potential \cite{m2}.

This paper is concerned with two fundamental issues in such no-scale models of inflation,
the incorporation of supersymmetry breaking and the identification of the inflaton field.
Particle experiments hope to constrain the pattern of soft supersymmetry 
breaking, which is sensitive to the form of the effective supergravity theory, and
the breaking of supersymmetry alters the form of the effective inflationary potential,
in general. Scenarios for supersymmetry breaking within the minimal supersymmetric
extension of the Standard Model (MSSM) that are frequently studied include the constrained
MSSM (CMSSM) in which the soft supersymmetry-breaking scalar masses $m_0$, bilinear terms $B_0$,
trilinear terms $A_0$ and gaugino masses $m_{1/2}$
are universal at some input scale \cite{cmssm}. An interesting special case is that found in minimal
supergravity (mSUGRA), where $A_0 = B_0 + m_0$~\cite{bfs,vcmssm}. On the other hand, no-scale supergravity naturally leads
to the input conditions $m_0 = B_0 = A_0 = 0$ \cite{no-scale,LN}. Generating $m_{1/2} \ne 0$
requires a non-trivial gauge kinetic function in the effective supergravity theory, as may be generated in
the underlying string theory or through anomalies \cite{anom} 
as in the case of pure gravity mediation (PGM) \cite{pgm,eioy}. 
As we discuss in this paper, CMSSM, mSUGRA, no-scale, and PGM boundary conditions
may all be generated within the no-scale inflationary framework.

On the other hand, cosmological observations are providing ever tighter constraints on models of
inflation, via measurements of the tilt in the spectral index of scalar perturbations, $n_s$,
of the tensor-to-scalar perturbation ratio, $r$, and of non-Gaussian parameters such as
$f_{NL}$ \cite{planck15}. The measurements of $n_s$ and $r$, in particular, constrain the form of the
inflationary potential and the number of e-folds, providing information about the form
of the effective supergravity theory, the identification of the inflaton, and the couplings
that control its decays. As we shall see, these decays, and hence the reheating temperature
following inflation and the predicted values of $n_s$ and $r$ are sensitive not only to the
identification of the inflaton field but also to the mechanism and magnitude of supersymmetry breaking.

The purpose of this paper is to study the interplay of these cosmological and particle
constraints on the effective no-scale supergravity model of inflation arising from string theory,
showing how no-scale inflation may thereby serve as a bridge between string theory
and LHC physics.

Section~2 of this paper contains a brief review of relevant aspects of the no-scale
supergravity framework, and Section~3 introduces no-scale scenarios for inflation.
Possible patters of soft supersymmetry breaking within these scenarios are discussed in Section~4,
and inflaton decays and reheating are discussed in Section~5. Finally, Section~6
summarises our results and indicates possible directions for future work on no-scale inflation.

\section{Review of the No-Scale Supergravity Framework}

As was shown in~\cite{Witten1985}, generic string compactifications yield no-scale
supergravity as the effective field-theoretical framework for sub-Planckian physics.
In a large class of string compactifications, including orbifold examples~\cite{ADEH}, at the
lowest-genus level the K\"ahler potential $K$ for the dilaton and untwisted moduli fields has the general form
\beq
K \; = \; - \ln (S + {\bar S} ) - \sum_i \ln (T_i + {\bar T_i} ) - \sum_j \ln (U_j + {\bar U_j} ) \, ,
\label{KSTU}
\eeq
where $S$ is the dilaton, the first sum is over the $h_{1,1}$ untwisted K\"ahler moduli $T_i$, and
the second sum is over the $h_{2,1}$ untwisted complex structure moduli $U_j$. We recall that
the untwisted K\"ahler moduli parameterise the sizes of the compactification tori, and that the
complex structure moduli parametrise their complex deformations. In general, both $h_{1,1} \ge 3$
and $h_{2,1}$ are model-dependent: here we assume the minimum value $h_{1,1} = 3$. We
also assume that the dilaton $S$ and the complex structure moduli $U_j$ are fixed, as well as the
relative sizes of the untwisted K\"ahler moduli, so that we may simplify
\beq
- \ln (S + {\bar S} ) - \sum_i \ln (T_i + {\bar T_i} ) - \sum_j \ln (U_j + {\bar U_j} ) \; \to \; - 3 \ln (T + {\bar T} ) \, ,
\label{TitoT}
\eeq
where we term $T$ the volume modulus. Untwisted matter fields $\phi_\alpha$ may then be included via the substitution
\beq
T + {\bar T} \; \to \;  T + {\bar T} - \frac{1}{3} \sum_\alpha |\phi_\alpha|^2 \, .
\label{includeuntwistedmatter}
\eeq
Finally, we include in the lowest-genus effective K\"ahler potential twisted matter fields $\vphi_a$
with generic modular weights $n_a$, arriving at
\beq
K \; = \; - 3 \ln \left(T + {\bar T} - \frac{1}{3} \sum_\alpha |\phi_\alpha|^2\right) + \sum_a \frac{|\vphi_a|^2}{(T + {\bar T})^{n_a}} \, ,
\label{finalK}
\eeq
which we use as the basis of our subsequent discussion.

Equation (\ref{finalK}) is not the only possible starting-point for no-scale inflation and particle phenomenology,
since it embodies several assumptions about the moduli of the string compactification and
their stabilisation, but it is sufficiently general to have several relevant and interesting features,
as we explore in the subsequent Sections. 

\section{Scenarios for No-Scale Inflation}

Supersymmetry offers a natural framework for constructing inflationary models \cite{ENOT}
and, as the scales involved in these models approach the Planck scale, it is necessary
to consider these models in the context of supergravity \cite{nost,hrr,gl1}.
The simplest of such models in both simple \cite{hrr} and no-scale supergravity~\cite{EENOS}
make a very definite prediction for the scalar tilt in the microwave background anisotropy,
namely $n_s = 0.933$, which is now definitively excluded by Planck measurements \cite{planck15}. 
In contrast, the Starobinsky model of inflation based on a $R +R^2$ extension of gravity predicts 
$n_s = 0.965$ \cite{MC,Staro2}, in excellent agreement with the Planck value $n_s = 0.9653 \pm 0.0048$.

It was shown that no-scale supergravity could lead to a consistent $R +R^2$ extension of gravity~\cite{Cecotti},
and the Starobinsky model of inflation was derived recently from no-scale supergravity
models \cite{ENO6,ENO7}. Phenomenologically viable models of inflation in no-scale supergravity
generally require at least two chiral superfields. One of these fields is the volume modulus, $T$,
and the other an untwisted matter field, $\phi$.  In what follows, we will consider 
the phenomenological consequences of both possibilities when both untwisted and twisted matter
fields are added to the theory. In this context, we discuss supersymmetry breaking in the next Section,
and scenarios for inflaton decays and reheating in the following Section. As we shall see, these issues are
connected in non-trivial ways.

\subsection{No-Scale Inflationary Models and the Starobinsky Model}

In this simplest no-scale supergravity, the two complex fields, denoted here by $(T,\phi)$, parametrise
a $SU(2,1)/SU(2)\times U(1)$ coset space, and the K\"ahler potential may be written in the form
\beq\label{k_ns}
K \; = \; -3\ln\left(T+\bar{T}-\frac{|\phi|^2}{3} \right).
\eeq
We recall that, here and throughout the paper, we assume that the dilaton field $S$ has been
fixed by some unspecified dynamics. 
The effective Lagrangian stemming from this K\"ahler potential has the form
\beq
\mathcal{L} \; = \; (T+\bar{T}-|\phi|^2/3)^{-1} (\partial_{\mu}T,\partial_{\mu}\phi)\left(
\begin{matrix}
3 & -\phi\\
-\bar{\phi} & T+\bar{T}
\end{matrix}
\right) \left(
\begin{matrix}
\partial^{\mu}\bar{T}\\
\partial^{\mu}\bar{\phi}
\end{matrix}
\right) - \frac{\hat{V}}{(T+\bar{T}-|\phi|^2/3)^2} \, ,
\eeq
where
\beq
\hat{V}=|W^{\phi}|^2+\frac{1}{3}(T+\bar{T})|W^{T}|^2+\frac{1}{3}\left(W^{T}(\bar{\phi}\bar{W}_{\phi}-3\bar{W})+{\rm h.c.}\right).
\eeq
The kinetic terms and scalar potential are derived from
\beq
\begin{aligned}\label{skin_phi}
\mathcal{L}_{B,{\rm kin}}=G^I_J D_{\mu}\Phi_{I} D^{\mu}\bar{\Phi}^J \, ,
\end{aligned}
\eeq
and
\beq\label{spot_phi}
\mathcal{L}_{B,{\rm pot}} = - e^G (G_I (G^{-1})^I_JG^J-3) \, ,
\eeq
where the K\"ahler function $G$ is defined as
\beq
G = K + \ln |W|^2 \, ,
\label{defG}
\eeq
and first- (second-)order derivatives of $G$ with respect to generic fields [and their conjugates]
$\Phi_I [{\bar \Phi}^J]$ are denoted by $G^I [G_J]$ ($G^I_J$).
In (\ref{defG}) we denote by $W(T,\phi)$ the superpotential, and $W^{T} \equiv \partial W/\partial T$,
$W^{\phi} \equiv \partial W/\partial \phi$. Unless explicitly denoted, we will work in Planck units $M_P^2 = 1$, where $M_P^{-2} = 8\pi G_N$ refers to the normalized Planck mass.  We note that the scalar kinetic term is invariant with 
respect to the action of the $SU(2,1)$ group,
but the scalar potential is in general not invariant. This implies that, for a given superpotential, 
the roles played by $T$ and $\phi$ are in general not interchangeable. In particular, depending on the form of $W$, 
either $T$ or $\phi$ may play the role of the inflaton, and we consider both possibilities in this paper.

For example, it was found in \cite{ENO6} that the $T$-independent  Wess-Zumino superpotential
\beq\label{WZ_staro}
W=m\left(\frac{\phi^2}{2}-\frac{\phi^3}{3\sqrt{3}}\right) 
\eeq
leads to 
\beq
V = 3m^2{\rm sech}^2\left(\frac{\chi-\bar{\chi}}{\sqrt{3}}\right)\left |\sinh(\chi/\sqrt{3}) \left( \cosh(\chi/\sqrt{3})- \sinh(\chi/\sqrt{3}) \right) \right|^2 \, ,
\eeq
where $\chi=\sqrt{3}\tanh^{-1}(\phi/\sqrt{3})$.
 The potential for the normalised real part of the inflaton ($x \equiv \sqrt{2}$ Re$\chi$) now takes the form 
\beq
V = 3m^2 e^{-\sqrt{2/3}x} \sinh^2(x/\sqrt{6}) \, ,
\label{nswzpot}
\eeq
and is identical to 
the Starobinsky inflationary potential
\beq 
V =  \frac{3}{4} m^2 \left(1- e^{-\sqrt{2/3}x}\right)^2 \, ,
\label{r2pot}
\eeq
obtained from a higher derivative form of the gravitational action,
\beq
S=\frac{1}{2} \int d^4x \sqrt{-g} \left(R+\frac{R^2}{6m^2}\right) \, .
\eeq
This identification between the no-scale Wess-Zumino model and $R^2$ gravity is possible if the modulus $T$ has a fixed vacuum expectation value $\langle T\rangle =1/2$. For a generic expectation value $\langle T\rangle =c$, the superpotential is of the form $W=\tilde{m}\left(\phi^2/2-\phi^3/(3\sqrt{6c})\right) $, where $\tilde{m}=(2c)^{1/2}m$.

Conversely, with a superpotential of the form \cite{Cecotti}
\beq\label{tph_w}
W=\sqrt{3}m\phi(T-1/2),
\eeq
the modulus $T$ plays the role of the inflaton field, with a Starobinsky potential along the canonically-normalized real direction,
$T=\frac{1}{2}(e^{-\sqrt{2/3}t}+i\sqrt{2/3}\sigma)$, for $\phi$ fixed at zero.

Both forms (Eqs.~(\ref{WZ_staro}) and (\ref{tph_w})) can be generalized by making use of SU(2,1)
transformations, or by adding additional superpotential terms that do not affect the scalar potential
along the inflationary trajectory \cite{ENO7}. We make use of one such generalization below, 
based on the addition of terms such as 
\beq
\Delta W = \left[ \frac{(T-1/2)^n 2^{n-2} \phi}{(2T+1)^{n-2}} \right]
\eeq
that introduce new couplings between the inflaton ($\phi$ in this case) and the volume modulus.

\subsection{Symmetric Formulation}

The $SU(2,1)/SU(2)\times U(1)$ model can be rewritten equivalently in a more symmetric form with  K\"ahler potential
\beq\label{k_s}
K = -3\log\left(1-\frac{|y_1|^2+|y_2|^2}{3}\right) \, .
\eeq
In this basis, the $SU(2,1)$ transformations of the fields correspond to 
\beq
y_1\, \rightarrow\, \sqrt{3}\frac{A_{11}y_1 + A_{12}y_2 + \sqrt{3}A_{13}}{A_{31}y_1+A_{32}y_2+\sqrt{3}A_{33}} \ , \ \ y_2 \,\rightarrow\, \sqrt{3}\frac{A_{21}y_1 + A_{22}y_2 + \sqrt{3}A_{23}}{A_{31}y_1+A_{32}y_2+\sqrt{3}A_{33}} \ , 
\eeq
with $A\in SU(2,1)$. The complex fields $y_{1,2}$ are related to the fields $T,\phi$ by the relations
\beq
y_1 = \left(\frac{2\phi}{1+2T}\right)\quad , \qquad y_2=\sqrt{3}\left(\frac{1-2T}{1+2T}\right) \ ,
\label{YtoT}
\eeq
and their inverses
\beq\label{TtoY}
T=\frac{1}{2}\left(\frac{1-y_2/\sqrt{3}}{1+y_2/\sqrt{3}}\right) \quad , \qquad \phi = \left(\frac{y_1}{1+y_2/\sqrt{3}}\right).
\eeq
Simultaneously, the superpotential is transformed as
\beq\label{TtoY_W}
W(T,\phi)\, \rightarrow\, \widetilde{W}(y_1,y_2) = \left(1+y_2/\sqrt{3}\right)^3W\, .
\eeq
Unsurprisingly, under the transformations (\ref{TtoY}),(\ref{TtoY_W}), inflationary potentials in the $(T,\phi)$ basis are 
mapped into inflationary potentials in the $y_{1,2}$ basis \cite{ENO7}. In particular, the Wess-Zumino superpotential (\ref{WZ_staro}) transforms to
\beq
\widetilde{W} = m\left[\frac{y_1^2}{2}\left(1+\frac{y_2}{\sqrt{3}}\right)-\frac{y_1^3}{3\sqrt{3}}\right] , 
\eeq
for which the Starobinsky potential is recovered along the canonically normalized $y_1$ direction for $y_2=0$. 
Analogously, the superpotential (\ref{tph_w}) transforms into
\beq
\widetilde{W} = m y_1 y_2 \left(1+y_2/\sqrt{3}\right)\, ,
\eeq
where now $y_2$ plays the role of the inflaton, with a Starobinsky potential along its real direction.

\subsection{Incorporation of Twisted Matter}

In~\cite{EGNO3,EGNO2} we found a different realization of inflation within a no-scale setting,
by considering a no-scale structure with a twisted matter field with a sum of modular weights $\sum_i n^i_a = 3$,
described by the K\"ahler potential
\beq
K \; = \; - \sum_{i=1}^3 \ln(T_i + \bar{T}^i) + \frac{|\vphi|^2}{\prod_{i=1}^3(T_i+\bar{T}^i)}\ .
\eeq
In particular, when the ratios of the moduli are fixed at a high scale as in (\ref{includeuntwistedmatter}),
the K\"ahler potential can be written in the form
\beq\label{k_egno2}
K \; = \; -3\ln(T+\bar{T})+\frac{|\vphi|^2}{(T+\bar{T})^3} \, .
\eeq
Making the choice of superpotential
\beq\label{tvph_w}
W=\sqrt{3}m\vphi(T-1/2) \, ,
\eeq
it was shown that at $\vphi=0$ the effective potential for $T$ is sufficiently flat to allow inflation for any initial
condition in the complex $T$ plane far away enough from the origin. Moreover, the standard Starobinsky potential is recovered
along the (canonically-normalized) real direction, whereas the chaotic quadratic potential appears along the imaginary direction\footnote{See \cite{FeKR} for other attempts at quadratic chaotic inflation in a complexification of the
Starobinsky model.}.

\subsection{Phenomenological Issues}

Our goal in this work is to embed the inflationary models with K\"ahler potentials (\ref{k_ns}, \ref{k_egno2}) 
in a more complete supergravity model, including matter fields and a source of supersymmetry breaking.
As was pointed out in~\cite{EGNO2}, the addition of a {\it supersymmetry-breaking} sector modifies in 
general the form of the inflationary potential, and it is of interest to determine to what extent the 
conclusions drawn in the pure $(T,\phi)$ or $(T,\vphi)$ scenarios still hold~\footnote{Throughout this paper, 
$\phi$ will refer to an untwisted field and $\vphi$ will refer to a twisted field.}. Furthermore, 
previous studies \cite{ekoty,twyy} of {\it inflaton decays} in a no-scale set  up have shown that,
in the absence of a direct coupling of the inflaton to matter, or of a non-trivial gauge kinetic function,
the decays of the inflaton are completely suppressed at tree level.  It was assumed
in~\cite{ekoty} that the  K\"ahler potential possessed an overall no-scale $SU(N+1)/[SU(N)\times U(1)]$ symmetry,
including $N$ matter fields, with the volume modulus $T$ playing the role of the 
supersymmetry-breaking field with flat tree-level potential.
However, here we consider a generic no-scale model with the K\"ahler potential of the form (\ref{finalK}),
and we explore the phenomenological implications of this model for two different scenarios: one in which $T$ is the inflaton,
and another in which one of the untwisted matter fields $\phi_\alpha$ is responsible for inflation.
Matter fields may be either twisted or untwisted.

\section{Patterns of Supersymmetry Breaking}

In order to break supersymmetry, the superpotential must have a non-zero vacuum expectation value at the minimum of the scalar potential. We consider first scenarios in which the inflaton is identified with one of the untwisted matter fields $\phi_\alpha$,
which we denote by $\phi_1$. In such a case we know that a $T$-independent superpotential like (\ref{WZ_staro})
leads to an inflationary potential. The volume modulus $T$ is free to play a role in supersymmetry breaking, and
we discuss in this Section various options for achieving this while obtaining Starobinsky-like inflation
and zero vacuum energy.

\subsection{Scenarios with a Matter Inflaton}

\subsubsection{Supersymmetry Breaking via the Volume Modulus}

One possibility is to add a constant term to a superpotential that otherwise would have a vanishing 
vacuum expectation value (vev).
For definiteness we consider a generic superpotential of the form
\beq\label{w_phi}
\begin{aligned}
W&=W_{\rm inf}(T,\phi_1) + (T+c)^{\beta}W_2(\phi_i) + (T+c)^{\alpha}W_3(\phi_i) \\
&\qquad +(T+c)^{\sigma}W_2(\vphi_a) +(T+c)^{\rho}W_3(\vphi_a) +  \mu \, ,
\end{aligned}
\eeq
where $c$ is an arbitrary constant, and
$W_{2,3}$ denote bilinear and trilinear terms with modular weights that are in general non-zero.
If we assume vanishing $F$ terms for all the scalar fields: $\langle W^{I}\rangle = 0$, and vanishing vevs for all scalar fields
except $T$, the inflationary minimum $\phi_1=0$ corresponds to a supersymmetry-breaking minimum with
vanishing cosmological constant if the following constraints are satisfied,
\beq\label{const_cond}
\langle W^{TT}\rangle = \langle W^{T\phi_1}\rangle = 0 \, .
\eeq
These are trivially fulfilled for the Wess-Zumino superpotential (\ref{WZ_staro}). When $\{\phi,\vphi\}=0$,
the effective potential for $T$ is completely flat at the tree level, so the volume modulus has an undetermined vev, and
the gravitino mass
\beq
m_{3/2} = \frac{\mu}{(T+\bar{T})^{3/2}}
\eeq
varies with the value of the volume modulus.

The induced soft terms can readily be calculated\footnote{Related derivations of soft terms in string models with flux compactifications can be found in \cite{lnr}.}: 
they are sector-dependent and sensitive to the vev of $T$, and are given by 
\begin{align}\label{soft_T_gen}
\phi_\alpha: & \quad m_0=0\ , \quad B_0 = -\beta m_{3/2}\frac{(T+c)^{\beta-1}}{(T+\bar{T})^{1/2}} \ , \quad A_0=-\alpha m_{3/2}\frac{(T+c)^{\alpha-1}}{(T+\bar{T})^{1/2}} \, ,
\end{align}
\begin{align}\label{soft_T_genvphi}
\vphi_a: & \begin{dcases}
m_0=m_{3/2}\frac{(1-n_a)^{1/2}}{(T+\bar{T})^{n_a/2}} \, ,\\
B_0 = 2m_{3/2}\frac{(T+c)^{\sigma-1}}{(T+\bar{T})^{3/2}}\left[(1-n_a)(T+c)-\frac{\sigma}{2}(T+\bar{T})\right] \, ,\\
A_0 = 3m_{3/2}\frac{(T+c)^{\rho-1}}{(T+\bar{T})^{3/2}}\left[(1-n_a)(T+c)-\frac{\rho}{3}(T+\bar{T})\right] \, ,
\end{dcases}
\end{align}
One can immediately check that $G^{I}=0$ for $I=\phi_\alpha,\vphi_a$, and that $G^{T}=-3/(T+\bar{T})$.
Therefore, as expected, the Goldstino $\eta= \sum_l G^{I}\chi_{I}$ is the fermionic partner of the modulus $T$, namely the modulino $\chi_{T}$.


The previous results ignore the fact that one typically needs to fix the vacuum expectation value of the volume modulus $T$ during inflation.
In the case of the Wess-Zumino model (\ref{WZ_staro}), the Starobinsky potential is obtained for $\langle T\rangle =c$.
This vev may be fixed with the addition of strongly stabilizing terms in the K\"ahler potential of the form \cite{EKN,ENO7}
\beq\label{k_sstab}
K \; = \; -3\ln\left(T+\bar{T} + \frac{(T+\bar{T}-2c)^4 + d(T-\bar{T})^4}{\Lambda^2} - \frac{|\phi_1|^2}{3} + \cdots\right) + \cdots \, .
\eeq
This modification of $K$ fixes $T$ during inflation and generates a mass term for it.
If $\Lambda\ll 1 $, this mass is hierarchically larger than the gravitino mass:
\beq\label{t_mass}
m_{T}^2=144c(d+1)\frac{m_{3/2}^2}{\Lambda^2}\, . 
\eeq
With the addition of the stabilizing terms, the induced soft parameters (\ref{soft_T_gen}, \ref{soft_T_genvphi}) reduce to
\begin{align}\label{sft_mu_1}
\phi_i:& \quad m_0= 0\ , \qquad\qquad\qquad\,\ B_0= -\beta m_{3/2}\ , \qquad \qquad\quad\ \,\  A_0= -\alpha m_{3/2} \\ \label{sft_mu_2}
\vphi_a:& \quad m_0= (1-n_a)^{1/2}m_{3/2}, \ \ B_0= 2\left(1-n_a-\frac{\sigma}{2}\right)m_{3/2}, \ \ A_0= 3\left(1-n_a-\frac{\rho}{3}\right) m_{3/2}
\end{align}
after rescaling the fields, $\phi_i'=(2c)^{-1/2}\phi_i$, $\vphi_a'= (2c)^{-n_a/2}\vphi_a$,
where $\phi',\vphi'$ are canonically normalized, and upon rescaling:
\beq\label{Trescaling}
\begin{aligned}
&W_2(\phi_i)\rightarrow (2c)^{3/2-\beta}W_2(\phi_i')\ ,  &&W_3(\phi_i)\rightarrow (2c)^{3/2-\alpha}W_3(\phi_i')\,,\\
&W_2(\vphi_a)\rightarrow (2c)^{3/2-\sigma}W_2(\vphi_a')\ ,  &&W_3(\vphi_a)\rightarrow (2c)^{3/2-\rho}W_3(\vphi_a')\,
\end{aligned}
\eeq
with $m_{3/2}=(2c)^{-3/2}\mu$. The forms of the soft supersymmetry-breaking terms
for the twisted matter fields suggest that modular weights with $n_a>1$ in the K\"ahler potential are not consistent with this framework.
A careful analysis reveals that in such cases the fields evolve towards a global anti-de-Sitter (AdS) minimum. 

The forms of Eqs.~(\ref{sft_mu_1}) and (\ref{sft_mu_2}) open up various phenomenological possibilities, some of which we
now enumerate.

$\bullet$ If all matter fields are of the untwisted type, we see that there are no supersymmetry-breaking contributions
to scalar masses. If in addition, the modular weights $\alpha$ and $\beta$ vanish, then $A_0 = B_0 = 0$.
and we recover the original {\it no-scale} boundary conditions \cite{LN}.  Models with
radiative electroweak symmetry breaking \cite{ewsb} can be accommodated if these boundary conditions
are fixed at scales above the GUT scale \cite{pro,emo2,ENO8}. In addition, this yields a much more restrictive phenomenological 
parameter space than CMSSM-like models since the ratio
of the Higgs vevs, $\tan \beta$, is determined by the Higgs minimization conditions and is no longer a free parameter \cite{vcmssm}.

$\bullet$ However, if matter fields are of the twisted type,
then other possibilities arise. For simplicity, let us take the kinetic modular weights to be 0.
In this case, we have universal soft scalar masses as in {\it CMSSM}-like models, which
are determined by the gravitino mass \cite{bfs}. 

$\bullet$ On the other hand,
when the superpotential weights are equal ($\rho = \sigma$),
we obtain {\it mSUGRA}-like boundary conditions, with $A_0 = (3-\rho) m_{3/2}$ and $B_0 = (2-\rho) m_{3/2}$,
i.e., $B_0 = A_0 - m_0$~\cite{bfs,vcmssm}. These mSUGRA-like models also yield a much more restrictive phenomenological 
parameter space where, in the context of radiative electroweak symmetry breaking, the ratio
of the Higgs vevs, $\tan \beta$, is again determined by the Higgs minimization conditions and no longer a free parameter. 

$\bullet$ Had we chosen to work in the symmetric ($y_1,y_2$) basis with no superpotential
weights, we would find $\rho = \sigma = 3$, in which case $A_0 = 0$ and $B_0 = -m_{3/2}$.
If, in addition, there are no tree-level sources for gaugino masses, the models would
be equivalent to {\it pure gravity mediation} (PGM) with radiative electroweak symmetry breaking \cite{eioy}. 

$\bullet$ Finally,
we note that if the weights $n_a \ne 0$, we have a source for {\it non-universal scalar masses} in the twisted sector. 

Further examples of no-scale, CMSSM and mSUGRA patterns of soft supersymmetry breaking are
presented subsequently.

\begin{figure}[h!]
\centering
	\scalebox{0.5}{\includegraphics{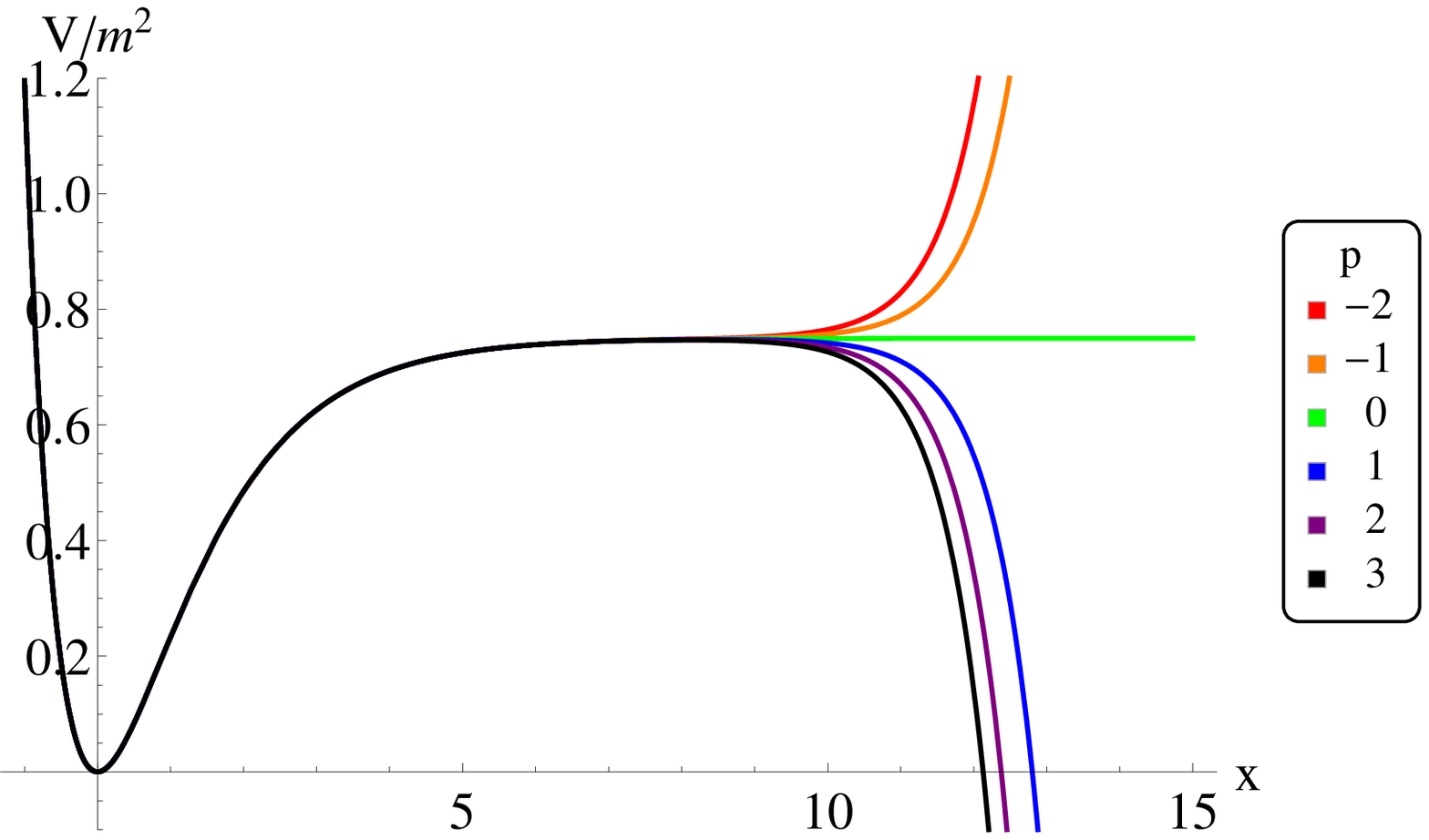}} \quad
	\scalebox{0.5}{\includegraphics{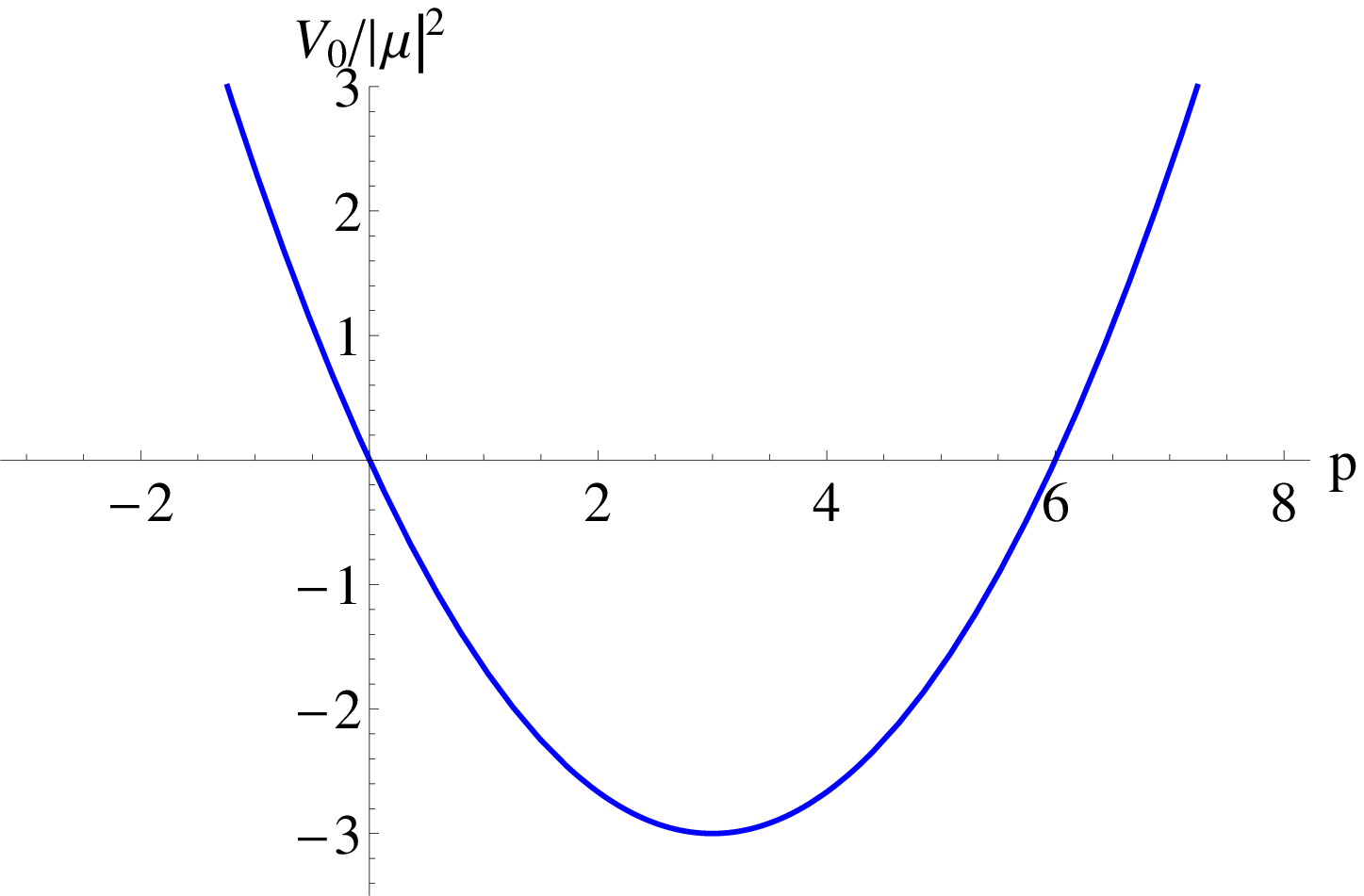}} \\
	\caption{\it Projections of the effective inflationary potential for the model (\ref{WZ_staro}) with the
	stabilised K\"ahler potential (\ref{k_sstab}) and the $T$-dependent superpotential $\Delta W=\mu(T+1/2)^{p}$,
	for different values of $p$, and $c=1/2$. Here $T$ is stabilized at $T=1/2$ with $\Lambda^{-2}=10$, and
	we use the nominal values $m=10^{-5}$, $\mu=10^{-13}$. 
	Left: The potential along the canonically-normalized real direction, $x=\sqrt{6}\tanh^{-1}(\phi_1/\sqrt{3})$.
	Right: The cosmological constant as a function of $p$.} 
	\label{mu_weight}
\end{figure}

One possible generalization of the superpotential (\ref{w_phi}) is to incorporate a modular weight for $\mu$:
\beq\label{mu_rep_1}
\mu \rightarrow \mu(T+c)^{p} \, .
\eeq
It is not difficult to verify that the scalar potential is not minimized at $(T,\phi_1)=(c,0)$ for generic
$p\neq 0$. However, with the addition of the stabilizing term (\ref{k_sstab}) one always has $\langle T\rangle\simeq c$
and a $p$-dependent inflationary potential of the form
\beq
\begin{aligned}
V= 3m^2 \cosh ^4\left(\frac{x}{\sqrt{6}}\right)& \left[\tanh ^4\left(\frac{x}{\sqrt{6}}\right)-2\tanh ^3\left(\frac{x}{\sqrt{6}}\right)\right.\\
&\quad\left.+ \left(1-  \frac{p\tilde{\mu}}{3m}\right)\tanh ^2\left(\frac{x}{\sqrt{6}}\right) +\left(\frac{p}{3} -2\right)   \frac{p\tilde{\mu}^2}{3m^2}\right]
\end{aligned}
\eeq
where
\beq
\tilde{\mu}=(2c)^{p-3/2}\mu,
\eeq 
and 
\beq\label{x_re}
x=\sqrt{6}\tanh^{-1}(\phi_1/\sqrt{6c})
\eeq 
denotes the canonically-normalized real part of $\phi_1$. The left panel of Fig.~\ref{mu_weight} shows the shape of the potential for
various values of $p$. As expected, for $p=0$ one exactly recovers the Starobinsky potential. When we assume a ratio $\mu/m = 10^{-8}$ (corresponding to $\mu \sim 100$ TeV) then
for $x\lesssim 9$ (during the inflationary phase) the potentials are almost indistinguishable, 
yielding a maximum of $N_{max}\simeq1160$ e-folds of inflation. For smaller $\mu$, the potential remains flat to higher $x$ and more e-folds of inflation are possible. However, the right panel of
Fig.~\ref{mu_weight} shows that there is, in general, a non-vanishing cosmological constant at $x=0$, where 
\beq
V_0 = \frac{|\tilde{\mu}|^2}{3}p(p-6) \, .
\eeq
Note that for $p >0$, the potential is unbounded from below and for $p<0$, we have a positive cosmological
constant, so that $p=0$ is the only possible solution in this case.

\subsubsection{Supersymmetry Breaking via the Polonyi mechanism}

As another possible generalization of the above set of models,  we can promote $\mu$ in (\ref{w_phi}) to have the form of the Polonyi superpotential \cite{pol},
dependent on a singlet field $z$:
\beq\label{mu_pol}
\mu \rightarrow \mu(z+\nu) \, ,
\eeq
which belongs to the twisted matter sector and has zero modular weight, assuming a strongly-stabilized K\"ahler potential \cite{dine,ego}
\beq\label{K_pol}
K \supset z\bar{z}-\frac{(z\bar{z})^2}{\Lambda_{z}^2}
\eeq
which might be due, e.g., to non-perturbative effects.
In the standard scenario, the second term in (\ref{K_pol}) drives the field $z$ to a supersymmetry breaking minimum
located at $z\simeq \Lambda_z^2/\sqrt{12}$, with the parameter $\nu\simeq 1/\sqrt{3}$ tuned to yield a 
vanishing cosmological constant \cite{dlmmo,ego}. In the present case, with a no-scale inflationary sector where $\phi_1$
is identified as the inflaton, the same values of $z,\nu$ with $T=c$, $\phi_1=0$ minimize the scalar potential.
However, this point in field space corresponds to a deSitter minimum with cosmological constant $V_0\simeq |\tilde{\mu}|^2$.

\begin{figure}[h!]
\centering
	\scalebox{0.5}{\includegraphics{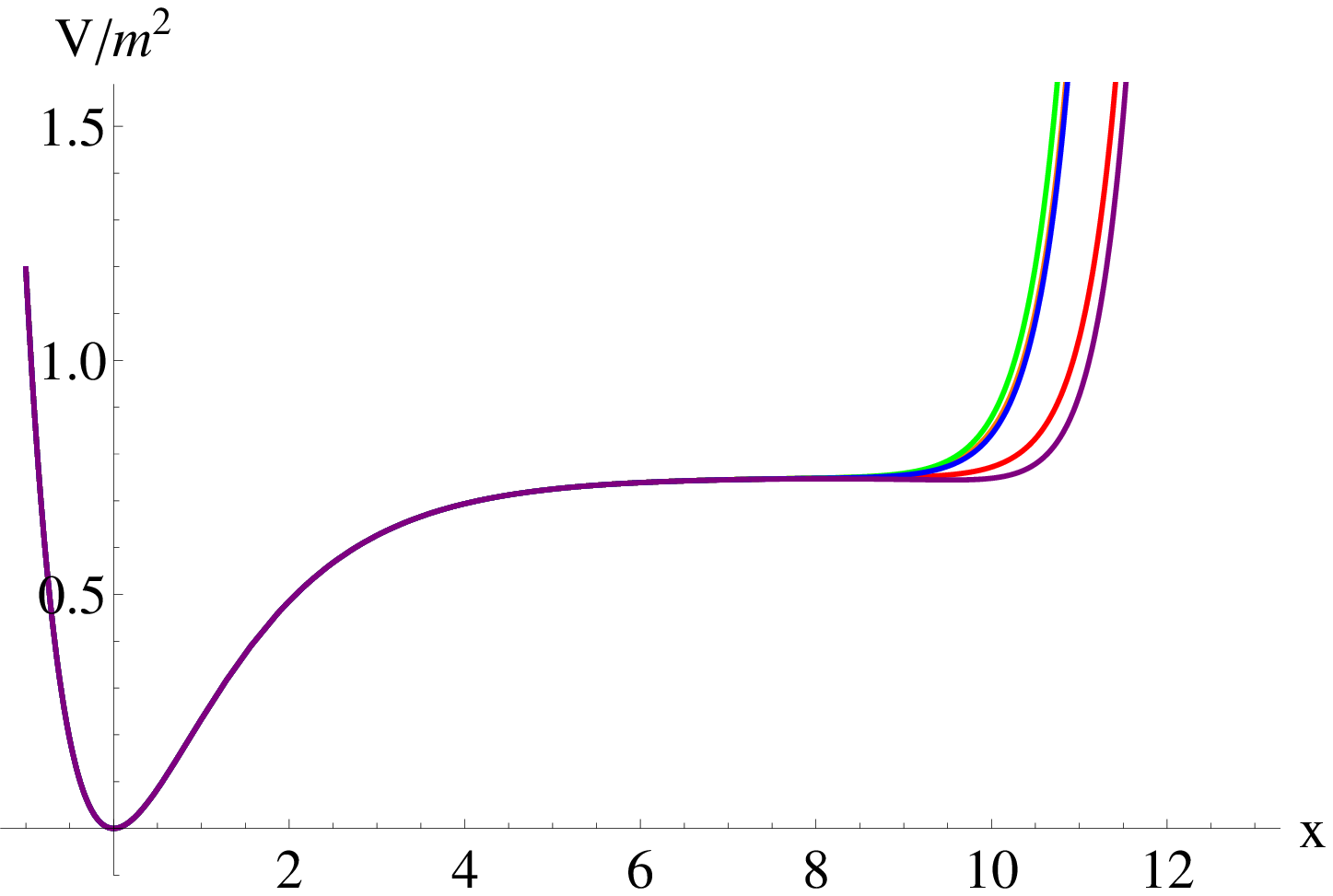}} \quad
	\scalebox{0.5}{\includegraphics{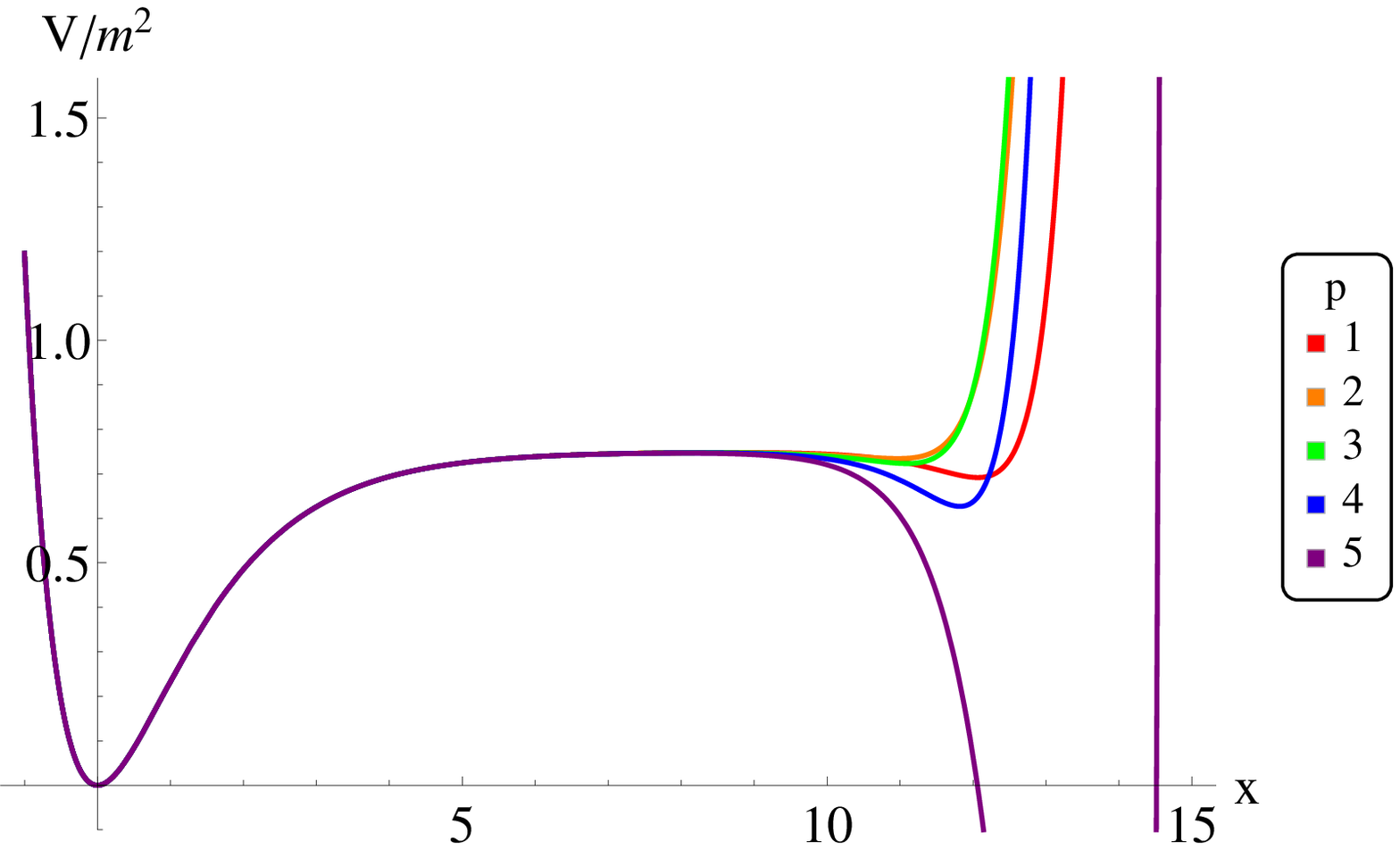}} \\
	\caption{\it Projections of the effective inflationary potential for the model (\ref{WZ_staro}) with the
	Polonyi sector (\ref{K_pol}) with superpotential $\Delta W=\mu(z+\nu)(T+1/2)^{p}$ (\ref{mu_pol_2}),
	for different values of $p$ and $c=1/2$. Here $T=1/2$, $z$ and $\nu$ are given by (\ref{znu}), and
	we use the nominal values $m=10^{-5}$, $\mu=10^{-13}$. 
	Left: The potential along the canonically-normalized real direction, $x=\sqrt{6}\tanh^{-1}(\phi_1/\sqrt{3})$,
	for $\Lambda_z=10^{-2}$.
	Right: Idem. for $\Lambda_z=4\times10^{-3}$.} 
	\label{pol_lam}
\end{figure}

This positive vacuum energy may be used to uplift a potential that would otherwise have an
AdS minimum. In particular, in the case of (\ref{mu_rep_1}) with $0<p< 6$, the extremum at $x=0$ can be uplifted if, 
instead of the superpotential (\ref{mu_rep_1}), we assume
\beq\label{mu_pol_2}
\mu \rightarrow \mu(z+\nu)(T+c)^{p}\,.
\eeq
The scalar potential is minimized with a zero cosmological constant at $(T,\phi_1)=(c
,0)$ for
\beq\label{znu}
z\simeq -\frac{(p^2 - 6p + 3)\Lambda_z^2}{4(3p(6-p))^{1/2}}\ , \qquad \nu\simeq \left(\frac{3}{p(6-p)}\right)^{1/2} \, ,
\eeq
assuming $\Lambda_z\ll 1$. For a small stabilizing parameter $\Lambda_z\ll \mu/m$, 
the superpotential (\ref{mu_pol_2}) becomes virtually indistinguishable from (\ref{mu_rep_1}). 
However, for the range of $p$ that we consider, the global minimum is not located at $x=0$, 
but corresponds to an AdS minimum located at
\beq
x_{\rm AdS}\simeq -\frac{1}{2}\sqrt{\frac{3}{2}}\log\left[\frac{3(6-p)\mu^2}{64m^2p}\right]\ , \qquad V_{\rm AdS}\simeq -\frac{4m^3}{3\sqrt{3}\tilde{\mu}}\left(\frac{p}{6-p}\right)^{3/2} \, .
\eeq
For larger, but still small values of $\Lambda_z$,  there is still a minimum at large $x$, but it is no
longer a global minimum. 
In order to avoid this minimum altogether, the stabilizing parameter $\Lambda_z$ must satisfy the constraint
\beq
\Lambda_z \gtrsim f(p)\left(\frac{\tilde{\mu}}{m}\right)^{0.3}\,,
\eeq
where $f\sim 2$ for $0<p\leq 4$, $f\simeq 4$ for $p=5$. This constraint is illustrated in Fig.~\ref{pol_lam} for $\mu/m=10^{-8}$.

%

The gravitino mass in this model is
\beq
m_{3/2}\; = \; \tilde{\mu}\left(\frac{3}{p(6-p)}\right)^{1/2} \, .
\label{Polonyigravitinom}
\eeq
Assuming a superpotential for the matter fields of the form (\ref{w_phi}), the induced soft parameters take the forms
\begin{align}\label{lambda_split}
\phi_i: & \begin{dcases}
m_0=\frac{1}{3}((6-p)p)^{1/2} m_{3/2} \, ,\\
 B_0= -\frac{1}{3}(p-\beta(p-3)) m_{3/2} \, ,\\
A_0 = \frac{1}{3}\alpha(p-3) m_{3/2} \, ,
\end{dcases}\\
\vphi_a: & \begin{dcases}
m_0=\frac{1}{3}(9-n_a(p-3)^{2} )^{1/2} m_{3/2} \, ,\\
 B_0= \frac{1}{3}(6+2n_a(p-3)+p(\sigma-3)-3\sigma) m_{3/2} \, ,\\
A_0 = -\frac{1}{3}(3-3n_a-\rho)(p-3) m_{3/2} \, .
\end{dcases}
\end{align}
In this case, the untwisted matter sector has non-vanishing soft supersymmetry-breaking
masses for any $0<p<6$, which are of universal CMSSM type.
If $p=3$, one has mSUGRA boundary conditions $m_0 = m_{3/2}, B_0 = - m_0$ and $A_0 = 0$ in both the untwisted 
and twisted sectors. Since the twisted-sector
soft supersymmetry-breaking parameters are independent of the modular weights $n_a$ for $p = 3$,
all values of the weights are allowed in this case. For $n_a = 0$ or for $p=2,4$ and modular weight $n_a=9$,
the soft supersymmetry-breaking scalar masses $m_0 = 0$ in the twisted sector. 

It is natural to consider if an untwisted Polonyi sector can provide the necessary 
uplifting for the superpotential (\ref{mu_rep_1}). It can be shown that this uplifting can be achieved, 
but at the cost of spoiling the inflationary potential. For example, using a superpotential of the form
\beq
\mu \rightarrow \mu\left(\frac{z}{\sqrt{3}}+\nu\right)^q(T+c)^{p} \, ,
\eeq
it is possible to show that the 
resulting potential is either unbounded from below or possesses an AdS minimum for large values of $x$, 
the canonically-normalized real component of $\phi_1$. 
The addition of strong stabilization terms for $z$ in the \kahler potential does not alleviate these problems.

\subsubsection{Incorporating the Giudice-Masiero Mechanism}

The Giudice-Masiero (GM) mechanism \cite{gm} is a well-known extension of minimal supergravity
in which a term proportional to $H_1H_2$
is introduced into the K\"ahler potential, so as to avoid the explicit introduction of a term $\mu_{H}H_1H_2$
in the superpotential with a coefficient $\mu_H$ with scale similar to that of electroweak symmetry breaking. 
In our no-scale framework, there are several ways to
implement this mechanism, depending on the sector to which the Higgs superfields belong. If the Higgs belong to the 
twisted sector with modular weights $n_1,n_2$, then a generic GM term of the form
\beq\label{GM1}
\Delta K = (T+\bar{T})^{-n_{12}}\left(c_H(T+c)^{\gamma}H_1H_2 + {\rm h.c.}\right)
\eeq
induces the $\mu$-term and a soft $B$-term,
\beq
\begin{aligned}
\Delta \mu_{H} &= (1-\tilde{n}_{12})c_Hm_{3/2} \, ,\\ 
\Delta B\mu &= \left[(1-\tilde{n}_{12})(2-\tilde{\gamma}-\tilde{n}_1-\tilde{n}_2+\tilde{n}_{12}) + \frac{p}{3}\tilde{n}_{12}\right]c_H m_{3/2}^2 \, ,
\end{aligned}
\eeq
where a tilde denotes a rescaling by a factor of $(3-p)/3$, e.g. $\tilde{\gamma}=\frac{3-p}{3}\gamma$, and where we have rescaled $c_H\rightarrow (2c)^{n_{12}-\gamma-(n_1+n_2)/2}c_H$. In the case of minimal (no) coupling to $T$, these reduce to $\Delta \mu_{H} =c_Hm_{3/2}$, $\Delta B\mu = 2c_H m_{3/2}^2$.
When the Higgs fields belong to the untwisted sector, a GM term such as (\ref{GM1}) generates soft terms of the form
\beq
\Delta \mu_{H} = (1-\tilde{n}_{12})c_Hm_{3/2}\ , \qquad
\Delta B\mu = \left[(1-\tilde{n}_{12})(\tilde{n}_{12}-\tilde{\gamma}+p/3)+p/3\right]c_H m_{3/2}^2.
\eeq
Alternatively, one can consider scenarios in which the GM term resides inside the logarithm.
One of the possibilities is
\beq
K = -3\ln\left[T+\bar{T}-\frac{1}{3}\left(|H_1|^2 + |H_2|^2  + (T+\bar{T})^{-q}(c_H(T+c)^{\gamma}H_1H_2 + {\rm h.c.}) + \cdots\right)\right] 
\eeq
for which 
\beq
\Delta \mu_{H}  = (p/3-\tilde{q}) c_Hm_{3/2}\ , \qquad
\Delta B\mu = \left[-\tilde{q}(1+\tilde{q}-\tilde{\gamma}-p/3)+\frac{p}{3}(2-\tilde{\gamma})\right]c_Hm_{3/2}^2,
\eeq
with $c_H\rightarrow (2c)^{q-\gamma}c_H$.

\subsubsection{Formulation in the Symmetric Basis}

For completeness, let us relate the phenomenology in the basis (\ref{k_ns}), that we have used so far, to the 
phenomenology in the more symmetric basis (\ref{k_s}).  In addition to the transformations (\ref{YtoT}), (\ref{TtoY}),
one must obtain the transformation rules for the matter fields $\{\phi,\vphi\}$. Denoting the matter fields 
belonging to the $(y_1,y_2)$ basis with a tilde ($\tilde ~$), we find the relations
\beq
\phi_i = \frac{\tilde{\phi}_i}{1+y_2/\sqrt{3}}\ , \qquad \vphi_a = \frac{\tilde{\vphi}_a}{(1+y_2/\sqrt{3})^{n_a}}\ ,
\eeq
and their inverses
\beq
\tilde{\phi}_i = \frac{2\phi_i}{1+2T}\ , \qquad \tilde{\vphi_a} = \frac{\vphi_a}{(T+1/2)^{n_a}}\ .
\eeq
The K\"ahler potential (\ref{finalK}) is then equivalent to
\beq
K = -3\ln\left[1-\frac{1}{3}\left(|y_1|^2+|y_2|^2+\sum_i |\tilde{\phi}_i|^2\right)\right] + \sum_a\frac{|\tilde{\vphi}_a|^2}{(1+|y_2|^2/3)^{n_a}} \, ,
\eeq
and the superpotential (\ref{w_phi}) is mapped into
\beq
\begin{aligned}
\widetilde{W} &= \widetilde{W}_{\rm inf}(y_1,y_2) + (1+y_2/\sqrt{3})^{1-\beta}W_2(\tilde{\phi}_i) + (1+y_2/\sqrt{3})^{-\alpha}W_3(\tilde{\phi}_i)\\
&  \qquad + (1+y_2/\sqrt{3})^{3-2n_a-\sigma}W_2(\tilde{\vphi}_a)  + (1+y_2/\sqrt{3})^{3-3n_a-\rho}W_3(\tilde{\vphi}_a) + \mu(1+y_2/\sqrt{3})^3 \, ,
\end{aligned}
\eeq
which leads to the same form of the soft supersymmetry breaking parameters given in (\ref{sft_mu_1}), as expected.
Analogous results hold for the generalizations of the parameter $\mu$ considered above, 
including the addition of a Polonyi sector, always recalling that upon changing basis one must substitute $\mu\rightarrow \mu(1+y_2/\sqrt{3})^3$.

\subsection{Scenarios in which the Volume Modulus $T$ is the inflaton}
\label{sec:T_inf}

It is also possible to identify the volume modulus $T$ with the inflaton. As we discussed before, a superpotential such as (\ref{tph_w}), 
which couples $T$ with a matter field $\phi$ (identified for simplicity with $\phi_1$) leads to a Starobinsky-like inflationary potential. While the simple form for supersymmetry breaking by a constant in $W$ does not
work in this case, we will see that the Polonyi mechanism and its generalizations will allow for 
successful phenomenological models. In the next subsubsection, we consider $\phi_1$ to be an untwisted field, and subsequently we will consider it to be a twisted field (labeled accordingly as $\varphi_1$).

\subsubsection{Inflation via Coupling to Untwisted Matter Fields}

For definiteness, we assume that the scalar fields $\{\phi,\vphi\}$ have vanishing vevs,
and we can consider the same superpotential (\ref{w_phi}) used in the previous subsection.
However, the conditions for a supersymmetry breaking minimum with vanishing cosmological constant (\ref{const_cond}) are not satisfied by the example (\ref{tph_w}) for $T$ field inflation. In fact, none of the multiple examples discussed in \cite{ENO7}
that yield inflationary potentials for $T$ satisfy these constraints. 

For example, when the constant term to the superpotential (\ref{w_phi}) is used as the source of supersymmetry breaking with the inflationary superpotential given by (\ref{tph_w}),
the minimum of the scalar potential is found at
\beq\label{Tinfmu}
T=\frac{1}{2}-\frac{\mu^2}{m^2} \ , \quad \phi_1 = \sqrt{3}\frac{\mu}{m}\ ,
\eeq
with cosmological constant $V_0=-3\langle e^{G}\rangle = -3m^2\mu^2/(m^2-3\mu^2)<0$.

This ADS vacuum must be lifted and we first attempt to use the 
a stabilized Polonyi sector as the source of supersymmetry breaking. With the K\"ahler potential (\ref{K_pol}) and superpotential (\ref{mu_pol}).
The supersymmetry breaking minimum is found at 
\beq
T\simeq \frac{1}{2} + \frac{2}{3}\left(\frac{\mu}{m}\right)^2\ , \ \ \phi_1\simeq \frac{\mu}{m}\ , \ \ z \simeq \frac{\Lambda_z^2}{\sqrt{12}}\ , \ \ \nu\simeq\frac{1}{\sqrt{3}}\left(1-\left(\frac{\mu}{m}\right)^2\right) \, ,
\eeq
for $\Lambda_z,\mu/m\ll1$. In this case, the form of the inflationary potential is unmodified from the Starobinsky form, save for the horizontal shift of the position of the minimum given by $t_0=-\sqrt{2/3}(\mu/m)^2$, where
\beq
T=\frac{1}{2}\left(e^{-\sqrt{\frac{2}{3}}t}+i\sqrt{\frac{2}{3}}\sigma\right) \, ,
\eeq
and $t$ denotes the canonically-normalized real part of $T$, which we associate with the inflaton. 
The supersymmetry breaking scale given by the gravitino mass is given by $m_{3/2}=\mu/\sqrt{3}$. The Goldstino in this case is identified with the fermionic partner of the Polonyi field, $\chi_z$, plus a small admixture of the fermion component of the $\phi_1$ superfield, $\chi_1$,
\beq
\eta \simeq \sqrt{3}\left(1-\left(\frac{\mu}{m}\right)^2\right)\,\chi_{z} + 3\frac{\mu}{m}\,\chi_1\,.
\eeq
When used with the superpotential (\ref{w_phi}) including matter 
we obtain the following universal soft parameters
\beq\label{soft_pol}
m_0=m_{3/2}\,\qquad B_0=-m_{3/2}\,\qquad A_0=0\, ,
\eeq
which are of the mSUGRA type when the gaugino masses are of order $m_{3/2}$ and of the PGM type if gaugino masses are generated through anomalies. Unlike the case where $\phi_1$ played the role of the inflaton (\ref{lambda_split}), we find no dependence for the soft parameters on the modular weights in (\ref{w_phi}). This is because, 
in general, the weight-dependent parts of the induced soft parameters are generated by the presence of the term $\langle (K^{-1})^T_TD^TW\rangle W^T$ in the effective scalar potential for the matter fields. In the present case of $T$ inflation, $\langle D^T W\rangle = \langle K^TW+W^T\rangle \propto \langle G^T\rangle=0$, and no contribution is generated. However,
different phenomenological boundary conditions could arise in the presence of $z$-dependent weights
for which $\langle G^z\rangle\neq 0$.

Supersymmetry can be broken by an untwisted Polonyi field that does not require stabilization in the K\"ahler potential if we choose
\beq\label{w_susy_1}
W_{\text{\cancel{susy}}} = \mu\left(\nu+\frac{z}{\sqrt{3}}\right)^3 \, . 
\eeq
As usual, the parameter $\nu$ must be tuned in order to have a vanishing cosmological constant. 
For $\mu\ll m$, the minimum can be approximately found to second order in $\mu/m$,
and is located at
\beq
T\simeq \frac{1}{2} + \left(\frac{\mu}{m}\right)^2\ , \ \ \phi_1\simeq \sqrt{3}\frac{\mu}{m}\ , \ \ z \simeq -\sqrt{3}\left(\frac{\mu}{m}\right)^2\ , \ \ \nu=1 \, .
\eeq
The shape of the inflationary potential is again unchanged from the Starobinsky form, except for a small shift of the position of the minimum, which corresponds now to $t_0=-\sqrt{6}(\mu/m)^2$.
Supersymmetry is broken by the non-vanishing vev of the superpotential (\ref{w_susy_1}), with the gravitino mass given by $m_{3/2}=\mu$. It is straightforward to check that $G^{I}\neq 0$ for $I=T,\phi_{1,2}$ if $\mu\neq0$. The Goldstino consists of $\chi_z$, with a small admixture of the modulino $\chi_T$, and $\chi_1$,
\beq
\eta \simeq \sqrt{3}\left(1-3\left(\frac{\mu}{m}\right)^2\right)\,\chi_2 + 2\sqrt{3}\frac{\mu}{m}\,\chi_1 + 3\left(\frac{\mu}{m}\right)^2\,\chi_T.
\eeq
The induced soft supersymmetry-breaking terms using (\ref{w_phi}) are given by 
\begin{align}\label{sft_mu_Tph1}
\phi_i:& \quad m_0= 0\ , \ \quad\qquad B_0= -m_{3/2}\,, \qquad A_0= 0\,, \\ \label{sft_mu_Tph2}
\vphi_a:& \quad m_0= m_{3/2}\,, \qquad B_0=  -m_{3/2}\,, \qquad A_0= 0\,.
\end{align}
In this case, the soft supersymmetry-breaking terms are universal in both sectors.
Since the soft supersymmetry-breaking
terms in the twisted sector are independent of the modular weights $n_a$,
the latter are unconstrained in this case. 
In the untwisted sector we have a special case of a mSUGRA-like spectra with $m_0 = A_0 = 0$, which would impose
universality above the GUT scale~\cite{emo}. In the twisted sector, we recover mSUGRA- or PGM-like models
with $m_0 = m_{3/2}$.

\subsubsection{Inflation via Coupling to Twisted Matter Fields}

The Starobinsky inflationary potential for the volume modulus $T$ can also be obtained by coupling $T$ with a matter field $\vphi$ with modular weight 3, belonging to the twisted matter sector. As we discussed before, a suitable superpotential has the form (\ref{tvph_w}). As before, we consider (\ref{w_phi}) as the form of the couplings to matter.  
The conditions for vanishing gradients and cosmological constant for the scalar potential at this point are completely analogous to the constraints (\ref{const_cond}) after replacing $\phi_1\rightarrow \vphi_1$ in $W_{\rm inf}$. Once again, these conditions are not compatible with $T$ field inflation using (\ref{tph_w}) or its avatars. Our example using a constant term in (\ref{w_phi}) displaces the minimum of the potential to
the approximate location given by (\ref{Tinfmu}) (with the replacement $\phi_1\rightarrow \vphi_1$).
It again corresponds to an AdS minimum, $V_0\simeq -3\mu^2$. We note that the solution in this case is only approximate due to the presence of the factor $e^{|\vphi_1|^2/(T+\bar{T})^3}$ in the scalar potential.

\begin{figure}[h!]
\centering
	\scalebox{0.5}{\includegraphics{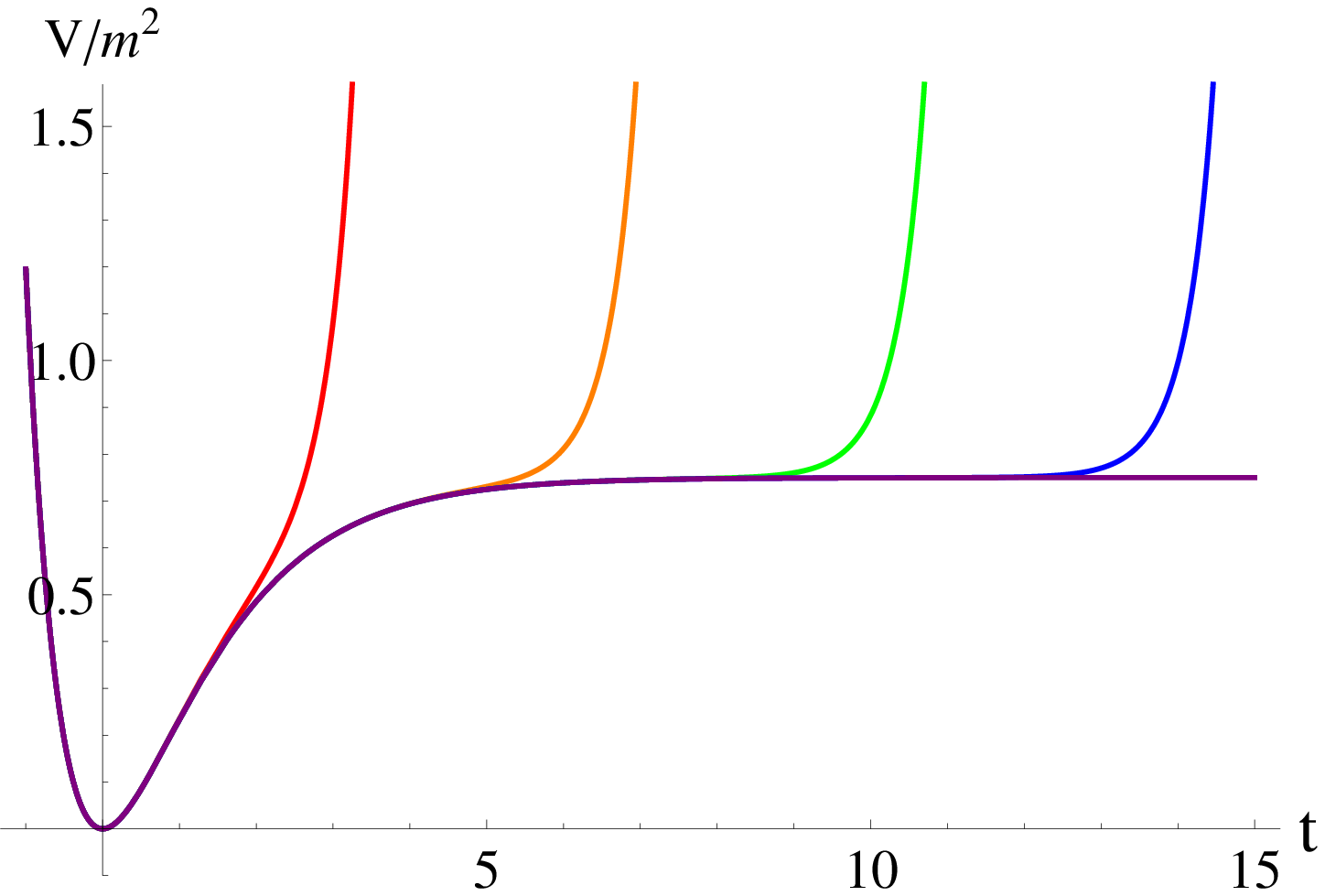}} \quad
	\scalebox{0.49}{\includegraphics{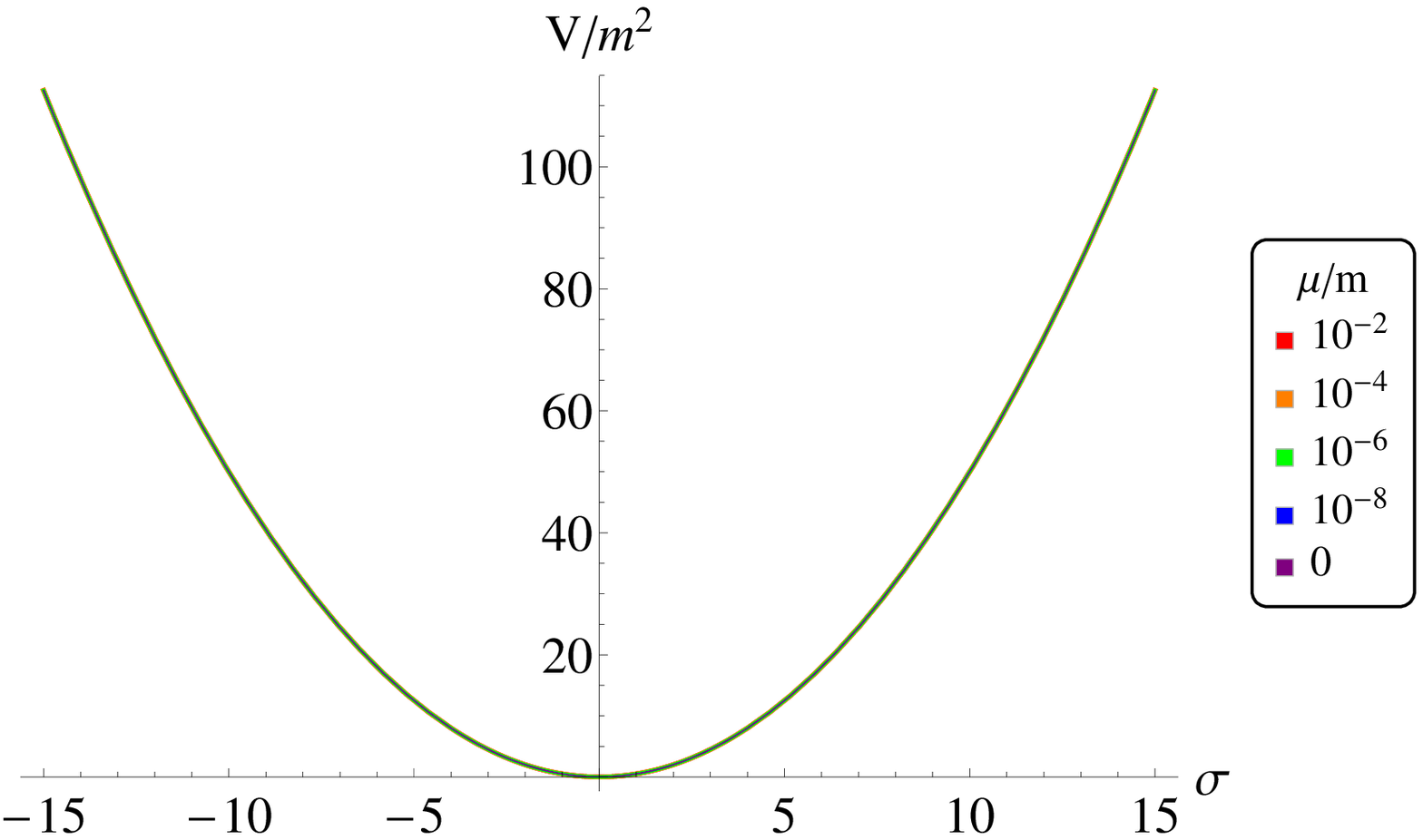}} \\
	\caption{\it Projections of the effective inflationary potential for the model (\ref{tvph_w}) with a Polonyi sector (\ref{mu_pol}), for different values of the ratio $\mu/m$. The fields $\vphi_1,\, z$ are assumed to have their minimum values, computed numerically.
	Left: The potential along the canonically normalized real direction, $t=-\sqrt{\frac{3}{2}}\log(2\,{\rm Re}T)$ .
	Right: The potential along the canonically normalized imaginary direction, $\sigma=\sqrt{6}\,{\rm Im}T$.} 
	\label{T_vphi_pol}
\end{figure}

Spontaneous supersymmetry breaking with a strongly stabilized Polonyi sector allows for a vanishing cosmological constant at the minimum given by
\beq
T\simeq \frac{1}{2} + \frac{2}{3}\left(\frac{\mu}{m}\right)^2\ , \ \ \vphi_1 \simeq \frac{\mu}{m}\ , \ \ z\simeq \frac{\Lambda_z^2}{\sqrt{12}}\ , \ \ \nu\simeq \frac{1}{\sqrt{3}}\left(1+\frac{1}{3}\left(\frac{\mu}{m}\right)^2\right)\,,
\eeq
for $\Lambda_z,\mu/m\ll1$. Unlike the $(T,\phi)$ scenario, this deformation is not limited to a shift in the position of the minimum; the behaviour of the potential at large values of the inflaton becomes dependent on the magnitude of $\mu$. In particular, along the real direction the potential receives the correction
 \beq
 \Delta V = \mu^2\left( e^{\sqrt{\frac{2}{3}}t} - 4e^{2\sqrt{\frac{2}{3}}t} + 3e^{\sqrt{6}t}\right)\,,
 \label{corr}
 \eeq
 (see Fig.~\ref{T_vphi_pol}), and the gravitino mass is $m_{3/2}=\mu/\sqrt{3}$. The Goldstino is a mixture of the fermion components of the $T$, $\vphi_1$ and $z$ superfields,
\beq
\eta \simeq \sqrt{3}\left(1-\frac{7}{3}\left(\frac{\mu}{m}\right)^2\right)\chi_z + 3\frac{\mu}{m}\,\tilde{\chi}_1 - 6\left(\frac{\mu}{m}\right)^2\chi_T.
\eeq
In this case the couplings to matter generate universal soft supersymmetry breaking terms of the mSUGRA type, given by
(\ref{soft_pol}).

Finally, we can also consider an untwisted Polonyi sector field with the cubic superpotential (\ref{w_susy_1}). In complete analogy to the scenario contemplated in the previous section, the non-vanishing vev of this superpotential shifts the position of the minimum of the potential. For $\mu\ll m$, this minimum is located at 
\beq
T\simeq \frac{1}{2} + \left(\frac{\mu}{m}\right)^2\ , \ \ \vphi_1\simeq \sqrt{3}\frac{\mu}{m}\ , \ \ z \simeq \frac{\sqrt{3}}{2}\left(\frac{\mu}{m}\right)^2\ , \ \ \nu=1 \, .
\eeq
The inflaton potential is again deformed by the addition of (\ref{w_susy_1}). Upon the addition of the Polonyi sector, the inflaton potential becomes dependent on $\mu$. To leading order in $\mu$, the potential correction along the real direction has the form  
 \beq
 \Delta V \simeq 3\mu^2\left( e^{\sqrt{\frac{2}{3}}t} - 3e^{2\sqrt{\frac{2}{3}}t} + 2e^{\sqrt{6}t}\right)\,.
 \eeq
 Although different from the correction in (\ref{corr}), the form of the potential looks very similar to that shown in Fig.~\ref{T_vphi_pol}.
%

In this case the gravitino mass given by $m_{3/2}=\mu$, and the Goldstino is
\beq
\eta \simeq \sqrt{3}\left(1-3\left(\frac{\mu}{m}\right)^2\right)\,\chi_2 + 2\sqrt{3}\frac{\mu}{m}\,\tilde{\chi}_1 - \frac{33}{2}\left(\frac{\mu}{m}\right)^2\,\chi_T.
\eeq
The induced soft parameters can be readily evaluated, and correspond to the CMSSM and mSUGRA
forms (\ref{sft_mu_Tph1}), (\ref{sft_mu_Tph2}) in the untwisted and twisted sectors, respectively.

\section{Inflaton Decays}

Any complete model of cosmological inflation should include mechanisms for inflaton decay
that yield successful reheating at the end of the inflationary epoch. In this Section we consider
inflaton decay in the model scenarios discussed in the previous Section, emphasising
differences in their corresponding predictions for the reheating temperature. These have important
phenomenological impacts, e.g., the inferred number of e-folds during inflation, the resultant
gravitino abundance and hence the possible scale of supersymmetry breaking, which may be
used to discriminate between models.

\subsection{Decay of the Untwisted Matter Inflaton}

We first calculate inflaton decays in
the scenario where the untwisted matter field $\phi_1$ plays the role of the inflaton,
assuming that all matter fields $\{\phi, \vphi\}$ have vanishing vevs at the end of inflation. This implies that
\beq\label{vev_cond}
\langle W^i\rangle = \langle W^a\rangle = 0 \ , \quad \langle K^i\rangle = \langle K^a\rangle =0 \, ,
\eeq
or, in terms of the \kahler function $G=K+\log|W|^2$,
\beq
G^{i}=G^a=0.
\eeq
The volume modulus $T$ must typically be stabilized in order to inflate successfully along the $\phi_1$
direction. As we have already seen, sufficient stabilization can be achieved by the addition of
quartic terms in the \kahler potential as in (\ref{k_sstab}). Thus, we assume now
that $T$ has a non-vanishing vev, $\langle T\rangle =1/2$, which implies $\langle G^T\rangle=p-3$ for the supersymmetry breaking superpotentials (\ref{mu_rep_1}), (\ref{mu_pol_2}), and the simple scenario with breaking by a constant (\ref{w_phi}), for which $p\equiv0$. In this case, 
all the matter scalar and fermion fluctuations about the global minimum are canonically normalized,
whereas the canonically-normalised modulus fluctuation corresponds to $\delta T=\sqrt{3}(T-1/2)$. 
For convenience, we define the ratio of the gravitino mass to the inflaton mass, $\Delta \equiv m_{3/2}/m$.

For the present analysis we consider a generic superpotential of the form
\beq
W=W_{\rm inf}(T,\phi_1) + W_{\rm M}(T,\phi_i,\vphi_a;\mu) \, ,
\eeq
for which we assume that the constraints (\ref{const_cond}) are satisfied. 
Here $\mu$ denotes the mass parameter that determines the scale of supersymmetry breaking:
$\langle W_{\rm M}\rangle=\mu$. A particular example corresponds to the superpotential 
(\ref{w_phi}) with $W_{\rm inf}(T,\phi_1)$ given by the Wess-Zumino superpotential (\ref{WZ_staro}). 
The decay rate of the inflaton is determined by its coupling to the moduli, matter and gauge fields. 
These couplings can be computed from a series expansion of the supergravity Lagrangian. For readability,
in the following discussion we drop the subscript $M$ from the matter superpotential, except when otherwise stated.

\subsubsection{Decays to matter scalars}

The interactions between the inflaton $\phi_1$ and the rest of the matter sector are determined from the 
scalar kinetic and potential terms in the Lagrangian. The scalar kinetic term is given by (\ref{skin_phi}).
%
After substitution of the matter field and modulus vevs, the scalar kinetic term yields
no interaction terms relevant for the kinematically-allowed decays up to four-body interactions.
We therefore look at interactions stemming from the potential term in the Lagrangian. 
Recall that the gauge-independent part of the scalar potential is given in (\ref{spot_phi})
and can be expanded to find the decay couplings. 
It is straightforward to calculate the scalar mass matrix, 
which takes the form
\beq\label{mass_m}
\bar{\Phi}^I(\mathcal{M}^2)^J_I\Phi_J = 
\left( \begin{matrix}
\bar{\phi}^1 & \bar{\Phi}^I
\end{matrix}
\right) \left(
\begin{matrix}
m^2 + m(W^{11}+\bar{W}_{11})+W^{1K}\bar{W}_{K1} & mW^{1J}+W^{JK}\bar{W}_{K1}\\
m\bar{W}_{1I}+W^{1K}\bar{W}_{KI} & W^{JK}\bar{W}_{KI}
\end{matrix}
\right) \left(
\begin{matrix}
\phi_1\\
\Phi_J
\end{matrix}
\right) \, ,
\eeq
where we denote $\Phi\equiv\{\delta T,\phi_i,\vphi_a\}$ and introduce the multiindex $I=\{\delta T,i,a\}$. 
Here we have segregated the inflaton explicitly from the rest of the matter and moduli fields,
and we have associated the inflaton mass $m$ with the vev of $W_{\rm inf}^{11}$, 
as is true for the Wess-Zumino superpotential (\ref{WZ_staro}).  It is immediately evident that, 
in the absence of a direct coupling between the inflaton and other fields in the matter superpotential, 
the field $\phi_1$ is the inflaton mass eigenstate~\footnote{We have ignored 
subdominant $\mathcal{O}(\mu)$ contributions in the expression (\ref{mass_m}),
which actually vanish for a $\phi_1$-independent matter superpotential.}.

A direct coupling between $\phi_1$ and the rest of the matter sector may be allowed. For example, 
this field can be associated with a heavy singlet sneutrino \cite{snu,ENO8}. In such case, one can consider  the addition of a Yukawa-like term
\beq\label{w_rhn}
\Delta W = y_{\nu} H_u L\phi_1
\eeq
to the Standard Model superpotential, where $y_{\nu}$ denotes the Yukawa coupling. Such a coupling leads to a mass matrix characteristic of seesaw models,
\beq
\left( \begin{matrix}
\bar{\phi}^1 & \bar{\tilde{\nu}}
\end{matrix}
\right) \left(
\begin{matrix}
m^2 + \tilde{m}^2 & -m\tilde{m} \\
-m\tilde{m} & \tilde{m}^2+\kappa\mu^2
\end{matrix}
\right) \left(
\begin{matrix}
\phi_1\\
\tilde{\nu}
\end{matrix}
\right) \, ,
\eeq
where $\tilde{m}\equiv y_{\nu} \langle H_u \rangle = y_{\nu} v \sin\beta$, and $\kappa=(1-n_{\nu})$ for a twisted neutrino, $\kappa=0$ for an untwisted neutrino. 
Therefore, even in the presence of direct couplings, we can consider $\phi_1$ to be
the inflaton mass eigenstate, up to corrections of order $\mu/m,v/m\ll 1$.

In order to determine the decay rate of the inflaton $\phi_1$, we must consider couplings beyond quadratic interactions.
Expansion of the scalar potential yields
\beq\label{L2_phi}
\begin{aligned}
\mathcal{L}_{B,{\rm pot}} =\  &\frac{2}{\sqrt{3}}m\bar{W}_{1J}\phi_1\delta T\bar{\Phi}^J  -\frac{1}{\sqrt{3}}B^1_J\phi_1\delta T\bar{\Phi}^J -\frac{1}{3\sqrt{3}}W^{1TT}_{\rm inf}\bar{W}_{T J}\phi_1\delta T\bar{\Phi}^J \\ 
& - \frac{c_{I\delta T}}{3}W^{1I}\bar{W}_{J T}\phi_1\Phi_I\bar{\Phi}^J  - W^{1IK}\bar{W}_{KJ}\phi_1\Phi_I\bar{\Phi}^J -\frac{1}{6}mW^{1TT}_{\rm inf}\phi_1\delta \bar{T}\delta \bar{T} \\
&+ \frac{2}{\sqrt{3}}m\bar{W}_{1J}\phi_1 \delta \bar{T}\bar{\Phi}^J - \frac{1}{\sqrt{3}}B^1_J\phi_1\delta \bar{T}\bar{\Phi}^J-\frac{1}{2}m\bar{W}_{1IJ}\phi_1\bar{\Phi}^I\bar{\Phi}^J \\
&- \frac{1}{2}W^{1K}\bar{W}_{KIJ}\phi_1\bar{\Phi}^I\bar{\Phi}^J - \frac{c_{IJ}}{6}W^{1T}\bar{W}_{IJ}\phi_1\bar{\Phi}^I\bar{\Phi}^J+ {\rm h.c.} + \mathcal{O}(\mu) + \cdots
\end{aligned}
\eeq
where we have introduced the notation
\beq\label{Bij}
B^{I_1I_2\dots}_{J_1J_2\dots}=\left[(n_a-3)W^{I_1I_2\dots a}\bar{W}_{aJ_1J_2\dots}-2W^{I_1I_2\dots k}\bar{W}_{kJ_1J_2\dots}\right]\,.
\eeq
and 
\beq
c_{IJ} = \left(
\begin{matrix}
-1 && -3 && n_J-2 \\
-3 && -5 && n_J-4\\
n_I-2 && n_I-4 && n_I+n_J-3
\end{matrix}
\right),
\eeq
where the rows and columns correspond to submatrices following the notation $I=\{\delta T,i,a\}$. The expression (\ref{L2_phi}) shows that all couplings to matter vanish in the absence of an explicit $\phi_1$
dependence in the matter superpotential, $W^{1I_1I_2\dots}=0$. It can be verified that the same is true for all the $\mathcal{O}(\mu)$ terms
that we have neglected in (\ref{L2_phi}), as well for any couplings leading to three- and four-body decay of the inflaton.
The only non-vanishing interaction in this limit correspond to those proportional to $W^{1TT}_{\rm inf}$. 
This coupling vanishes identically for the Wess-Zumino superpotential (\ref{WZ_staro}).
However, it is known that the superpotential (\ref{WZ_staro}) is not the unique superpotential that leads to 
Starobinsky inflation~\cite{ENO7}. Consider, e.g., the addition of the term
\beq\label{TTphi}
\Delta W_{\rm inf}= \zeta(T-1/2)^2\phi_1 \, ,
\eeq
which does not alter the shape of the potential for the inflaton Re~$\phi_1$ for any value of $\zeta$. 
In the presence of this term, the mass matrix has the structure 
\beq
m^2|\phi_1|^2 + m_T^2|\delta T|^2 + \frac{2\zeta}{3\sqrt{3}}(p-3)m_{3/2}M_P(\phi_1\delta T+ {\rm h.c.}) \, ,
\eeq
and the inflaton mass eigenstate corresponds to
\beq\label{tphieig}
\phi_1^M\simeq \phi_1 + (p-3)\frac{2\zeta \Delta M_{P}}{3\sqrt{3}m}\delta\bar{T} \, .
\eeq
In this case, the decay of the inflaton $\phi_1$ into the fluctuation of the modulus $T$ is possible, with rate
\beq\label{phitoT}
\Gamma(\phi_1 \rightarrow \delta T\,\delta T) = m \frac{|\zeta|^2}{72\pi}\, ,
\eeq
assuming that the modulus mass satisfies the hierarchy $m\gg m_{T}\gg m_{3/2}$ as in (\ref{t_mass}).
As we see in the next subsubsection, this is the same rate as the decay into gravitinos.
If these were the dominant decay rates, the Universe would become dominated by moduli and gravitinos,
forcing their masses to exceed 10 TeV in order to obtain a reheating temperature above 1 MeV, 
and hence suitable for nucleosynthesis. However, in this case, decays into neutralinos are
liable to yield a relic neutralino density that is
far too large. Thus we can not afford decays to moduli (and gravitinos) to be the dominant decay product.

Decay of the inflaton into matter becomes possible only if we allow a non-trivial dependence on $\phi_1$ for $W_M$. 
In particular, the superpotential (\ref{w_rhn}) leads to a non-vanishing amplitude
for which the dominant contribution corresponds to the seventh term in (\ref{L2_phi}) if $W^{1IJ} \ne 0$, namely
$-\frac{1}{2}m\bar{W}_{1IJ}\phi_1\bar{\Phi}^I\bar{\Phi}^J$. In the particular case of sneutrino inflation,
this coupling would be $-my_{\nu}\bar{H}_u\bar{\tilde{L}}\phi_1$, and
the decay width would be given by
\beq\label{phisneu}
\Gamma(\phi_1 \rightarrow H_u^0\tilde{\nu},H_u^+\tilde{f}_L) = m \frac{|y_{\nu}|^2}{16\pi}\, ,
\eeq
where we have neglected the masses of the final-state particles. This decay rate would be
fast if $|y_\nu | = {\cal O}(1)$ and, in order to avoid problems associated with gravitino production 
during reheating, we must set a bound on the Yukawa coupling associated with the inflaton~\cite{ENO8}
\beq
y_\nu \la 10^{-5} \, 
\eeq
with a corresponding constraint on the reheating temperature that we discuss below.

\subsubsection{Decays to matter fermions}

The decay of the inflaton $\phi_1$ to matter fermions is mediated by the interactions determined 
by the fermion kinetic term, the fermion mass matrix and the fermion-scalar interactions of the supergravity Lagrangian.
The fermion kinetic term is given by
\beq\label{fer_kin}
\mathcal{L}_{F,{\rm kin}} = \frac{i}{2}G^I_J\bar{\chi}_{IL}\gamma^{\mu}D_{\mu}\chi_{L}^J + {\rm h.c.} \, ,
\eeq
and yields no couplings relevant to two-, three- and four-body decays. 
One must then look for interactions stemming from the fermion mass matrix and the fermion-scalar interactions.
Working in the unitary gauge, one finds no dependence on the modulino $\chi_T$, 
which becomes the longitudinal component of the gravitino,
\begin{align}
\mathcal{L}_{F,{\rm int}} = &\ \frac{i}{2}\bar{\chi}_{IL}\slashed{D}\Phi_{J}\chi_{L}^K(-G^{IJ}_K+\frac{1}{2}G^{I}_K G^J) \notag \\ \label{fer_ints}
&\ +\frac{1}{2}e^{G/2}(-G^{IJ}-G^I G^J + G^{IJ}_K (G^{-1})^K_A G^A)\bar{\chi}_{IL}\chi_{JR} + {\rm h.c.}  \\
&\ + \text{four-fermion terms} \notag\\
 = &\ -\frac{1}{2}W^{1IJ}\phi_1\bar{\chi}_{I L}\chi_{J R} + \frac{i}{4\mu}W^{1J}\Phi_J\bar{\chi}_{KL}\slashed{\partial}\phi_1 \chi^K_L + \frac{i}{4\mu}W^{1J}\phi_1\bar{\chi}_{KL}\slashed{\partial}\Phi_J \chi^K_L \notag\\ \label{phi_fer_coup}
 &\ +\frac{1}{4\mu}W^{1J}W^{IK}\phi_1\Phi_J \bar{\chi}_{IL}\chi_{KR} - \frac{1}{2}W^{1IJK}\phi_1\Phi_J \bar{\chi}_{IL}\chi_{KR}\\
 &\ + \frac{\sqrt{3}}{2}W^{1JK}\phi_1({\rm Re}\,\delta T)\bar{\chi}_{JL}\chi_{KR} - \frac{1}{2}W^{1K}\phi_1\bar{\Phi}^J(\bar{\chi}_{KL}\chi_{JR} + \bar{\chi}_{JL}\chi_{KR}) + \cdots \notag
\end{align}
Similarly to the scalar case, all couplings to matter fermions vanish for a $\phi_1$-independent matter superpotential. 
The decay into a fermion and a higgsino is possible if we identify $\phi_1$ with a singlet neutrino, 
with superpotential (\ref{w_rhn}). In this case, the rate is given by 
\beq\label{phineu}
\Gamma(\phi_1\rightarrow \tilde{H}_u^0\nu,\,\tilde{H}_{u}^+f_L) =  m \frac{|y_{\nu}|^2}{16\pi} \, ,
\eeq
i.e., equal to the rate of decay into scalars. 

\subsubsection{Decay to the gravitino and inflatino}

We now explore the possibility of the decay of the inflaton $\phi_1$ to the gravitino. In the unitary gauge, 
this process is in general mediated by the interaction terms
\beq\label{gravitino_L}
\mathcal{L}_{3/2} =  \frac{1}{8}\epsilon^{\mu\nu\rho\sigma}\bar{\psi}_{\mu}\gamma_{\nu}\psi_{\rho}G^I\partial_{\sigma}\Phi_I + \frac{i}{2}e^{G/2}\bar{\psi}_{\mu L}\sigma^{\mu\nu}\psi_{\nu R} + {\rm h.c.} \, .
\eeq
Since $\langle G^I\rangle = 0$ for all matter fields, and $\langle G^T\rangle=p-3$, 
the couplings vanish unless there is mixing between the inflaton $\phi_1$ and the volume modulus $T$. 
Such mixing is possible in the presence of a term such as (\ref{TTphi}), in which case
the mass eigenstate is $\phi_1^M$, given by (\ref{tphieig}). In this case, the interaction is mediated by the Lagrangian
\beq
\mathcal{L}_{3/2} \simeq -\frac{\zeta m_{3/2}}{2m^2}\left[\frac{1}{2}\,\epsilon^{\mu\nu\rho\sigma} \bar{\psi}_{\mu}\gamma_{\nu}\psi_{\rho}\partial_{\sigma}\phi_1^M  -im_{3/2}\phi_1^M \,\bar{\psi}_{\mu}\sigma^{\mu\nu}\psi_{\nu}\right] \, .
\eeq
This results in the decay rate
\beq\label{phito32}
\Gamma(\phi_1 \rightarrow \psi_{3/2}\psi_{3/2}) \simeq m \frac{|\zeta|^2}{72\pi}.
\eeq
The same result is found for the decay of $\phi_1$ to the canonically-normalized modulino $\chi_{T}$,
the relevant coupling in this case being given by 
$\mathcal{L}\supset- \frac{1}{6}W^{1TT}_{\rm inf}\phi_1\bar{\chi}_{\delta T L}\chi_{\delta T R}$.

The decays to a single gravitino and a matter fermion are mediated by the interaction terms
\begin{align}\label{sing_grav}
\mathcal{L}_{3/2,\chi} &= \frac{i}{\sqrt{2}}e^{G/2}G^I \bar{\psi}_{\mu L}\gamma^{\mu}\chi_{I L} + \frac{1}{\sqrt{2}}G_J^I \bar{\psi}_{\mu L}\slashed{D}\bar{\Phi}^I\gamma^{\mu}\chi_{J R} + {\rm h.c.}\\
&= \frac{i}{\sqrt{2}}W^{1J}\phi_1 \bar{\psi}_{\mu L}\gamma^{\mu}\chi_{j L} + \frac{i}{\sqrt{2}}m\phi_1 \bar{\psi}_{\mu L}\gamma^{\mu}\chi_{1 L} + \frac{1}{\sqrt{2}}\bar{\chi}_{1R}\gamma^{\mu}\slashed{\partial}\phi_1\psi_{\mu L} + \cdots \, .
\label{sing_grav_2}
\end{align}
The decay amplitude to a matter fermion different from the inflatino is zero, 
unless there is an explicit dependence on $\phi_1$ in the matter superpotential. 
Identifying the inflaton with a singlet neutrino, with a coupling given by (\ref{w_rhn}), 
the decay to a left-handed neutrino and a single gravitino is allowed, but
with a negligible width relative to the decays to the Higgs, fermions and their supersymmetric partners:
\beq
\Gamma (\phi_1 \rightarrow \psi_{3/2}\,\nu) =v^2\sin^2\beta m \frac{|y_{\nu}|^2}{32\pi M_P^2} \sim 10^{-32}\, \Gamma(\phi_1\rightarrow \tilde{H}_u^0\nu,\,\tilde{H}_{u}^+f_L) \, .
\label{phigneu}
\eeq

Equation (\ref{sing_grav_2}) includes the interaction between the gravitino and the inflatino. 
The availability of this decay channel is strongly dependent on the mechanism of supersymmetry breaking. 
In the simplest scenario (\ref{w_phi}), the decay is not kinematically allowed, 
since there is no mass splitting at tree level for the untwisted matter field $\phi_1$. 
In the case when a splitting exists, such as (\ref{lambda_split}), the decay will be suppressed \cite{nos}
due to the degeneracy $m-m_{\chi}\sim m_{3/2}$:
\beq
\Gamma(\phi_1 \rightarrow \psi_{3/2}\,\chi_1) \sim \left(\frac{m_{3/2}}{m}\right)^2\frac{17 m^3}{48\pi M_P^2} \, .
\eeq
It can also be shown that all two-body decays involving one inflatino and one matter fermion $\chi_J$ 
are dependent on the coupling $W^{11J}$, which vanishes in the limit of no $\phi_1$ dependence in $W_M$, 
as well for a superpotential such as (\ref{w_rhn}).

We are led to conclude that, in the absence of a direct coupling between the inflaton 
and the rest of the matter (and gauge) sectors, there are no efficient decay channels for the inflaton,
if it is identified with an untwisted matter field, as found in other studies of no-scale supergravity \cite{ekoty}.
On the other hand, if the field $\phi_1$ is associated with a singlet neutrino, 
the decay rates (\ref{phisneu}) and (\ref{phineu}) imply a reheating temperature
\beq
T_R = (5.6\times 10^{14}\,{\rm GeV})|y_{\nu}|\left(\frac{g}{915/4}\right)^{-1/4}\left(\frac{m}{10^{-5}M_P}\right)^{1/2},
\eeq
assuming that the Yukawa coupling $y_{\nu}\lesssim\mathcal{O}(1)$ so that the decay of the inflaton
occurs after the end of inflation, during the oscillation of the inflaton around the minimum of the potential. Here $g$ denotes the effective number of degrees of freedom, and $g=915/4$ for the MSSM. 

\subsubsection{Decays to gauge bosons and gauginos}

The decay of the inflaton $\phi_1$ into gauge fields and gauginos is possible in the 
presence of a non-trivial coupling between $\phi_1$ and the gauge degrees of freedom,
as would be provided by a $\phi_1$-dependent gauge kinetic function 
$f_{\alpha\beta}=f(\phi_1)\delta_{\alpha\beta}$ \cite{ekoty,klor}. 
If supersymmetry is not broken by the inflaton,  this term will not contribute to gaugino masses.
These require a non-trivial dependence in the gauge kinetic function of fields involved in supersymmetry breaking. 
The relevant supergravity Lagrangian terms correspond to 
\beq\label{gaugeL}
\begin{aligned}
\mathcal{L}_G= -\frac{1}{4}&({\rm Re}\,f_{\alpha\beta})F_{\alpha\,\mu\nu}F_{\beta}^{\mu\nu} + \frac{i}{4}({\rm Im}\,f_{\alpha\beta})F_{\alpha\,\mu\nu}\tilde{F}_{\beta}^{\mu\nu}\\
& + \left( \frac{1}{4}e^{G/2}(\bar{f}_{\alpha\beta})_{,J}(G^{-1})^J_K G^K\bar{\lambda}_{\alpha L}\lambda_{\beta R} + {\rm h.c.}\right)
\, ,
\end{aligned}
\eeq
where $\tilde{F}_{\alpha}^{\mu\nu} = \frac{1}{2}\epsilon^{\mu\nu\rho\sigma}F_{\alpha\,\rho\sigma}$. 
Neglecting contributions suppressed by the gaugino masses, the decay widths
to canonically-normalized gauge boson pairs and gauginos can be evaluated in a straightforward way, resulting in \cite{ekoty}
\beq
\Gamma(\phi_1\rightarrow gg) = \Gamma(\phi_1 \rightarrow \tilde{g}\tilde{g}) = \frac{3d_{g,1}^2}{32\pi}\left(\frac{N_G}{12}\right)\frac{m^3}{M_P^2}\,,
\eeq
where $N_G$ is the number of final states: $N_G=12$ for the standard model, and $d_{g,1}$ is given by
\beq
d_{g,1} \equiv \langle {\rm Re}\,f\rangle^{-1}\left|\left\langle\frac{\partial f}{\partial \phi_1}\right\rangle\right|.
\eeq
The equality of the rates to gauge bosons and gauginos requires that $W_{\phi_1 \phi_1}$ is related to the inflaton mass rather than the supersymmetry-breaking scale. 
In the presence of a coupling such as (\ref{w_rhn}), these rates are subdominant, 
being suppressed by $(m/M_P)^2$ relative to the widths into Higgs, leptons and their supersymmetric partners,
cf, (\ref{phisneu}) and (\ref{phineu}). On the other hand, if no such couplings are present,
the decays to gauge bosons and gauginos are the dominant channels, and would yield a reheating temperature 
\beq
T_R =  (2\times 10^{10}\ {\rm GeV})\, d_{g,1}\,g^{-1/4}\left(\frac{N_G}{12}\right)^{1/2} \left(\frac{m}{10^{-5} M_P}\right)^{3/2} \, .
\label{phigg}
\eeq
In this case, the constraint on the thermal production of gravitinos is easily satisfied if
$d_g,1 \la 10^{-1}$. 

The decay of $\phi_1$ to gauge bosons and gauginos can also be achieved through a coupling 
between $T$ and the gauge degrees of freedom. Indeed, a $T$-dependent gauge kinetic function 
$f_{\alpha\beta}=f(T)\delta_{\alpha\beta}$ is a generic feature of heterotic string effective field theories~\cite{ADEH,gkf}. 
A superpotential such as (\ref{TTphi}) produces a mixing between $\phi_1$ and $T$, allowing
in this case decays of the $\phi_1$ mass eigenstate to gauge bosons, with a rate
\beq
\Gamma(\phi_1\rightarrow gg) = (p-3)^2\frac{d_{g,T}^2|\zeta|^2}{216\pi}\left(\frac{N_G}{12}\right)\Delta^2m\,,
\eeq
where we define
\beq\label{dgT}
d_{g,T} \equiv \langle {\rm Re}\,f\rangle^{-1}\left|\left\langle\frac{\partial f}{\partial T}\right\rangle\right| \, .
\eeq
We see, however, that this rate is suppressed by a factor $(m_{3/2}/m)^2$ relative to the decay widths 
(\ref{phitoT}) and (\ref{phito32}), and it can also be shown that the rate for decays
to gauginos is further suppressed by an additional $(m_{3/2}/m)^2$ factor.
Gaugino masses are generated in this case and are given by
\beq
m_{1/2} = \left| \frac{1}{2}e^{G/2}\frac{\bar{f}_{\alpha\beta,T}}{{\rm Re}\,f_{\alpha\beta}} (G^{-1})^T_T G^T \right| = \frac{d_{g,T}}{6}|p-3|m_{3/2}
\eeq
There is an additional contribution if $f_{\alpha\beta}$ also depends also on the Polonyi field $z$.

\subsection{Decays of a Volume Modulus Inflaton}

We will now consider the case where the inflaton is identified with the volume modulus $T$.
We assume as before that all matter fields $\{\phi,\vphi\}$ have vanishing vevs at the end of inflation,
which is equivalent to the conditions (\ref{vev_cond}). For all scenarios explored in Section~\ref{sec:T_inf}, 
the volume modulus inflaton $T$ has a non-vanishing vev at the minimum of the potential, which is located at
\beq
\langle {\rm Re}\,T\rangle=\frac{1}{2}+\mathcal{O}(\mu^2/m^2)\ , \quad \langle{\rm Im}\,T\rangle=0\,,
\eeq
where $\mu$ is the mass parameter that determines the scale of supersymmetry breaking:
$\mu=\sqrt{3}m_{3/2}$ for an untwisted Polonyi modulus $z$ with superpotential (\ref{mu_pol}), $\mu=m_{3/2}$ for breaking by an untwisted sector field with superpotential (\ref{w_susy_1}).
At this minimum we have $\langle W^T\rangle/\langle W\rangle=3+\mathcal{O}(\mu^2/m^2)$, and $\langle K^T\rangle =-3+\mathcal{O}(\mu^2/m^2)$. 
The decays of the inflaton are determined by its couplings to the moduli, matter and gauge fields,
which can be computed from a series expansion of the supergravity Lagrangian. In this subsection,
we denote $\Phi\equiv\{\phi,\vphi\}$ and use the multiindex $I=\{i,a\}$. 

\subsubsection{Decays to matter scalars}

The couplings of the inflaton $T$ to matter stem from the scalar kinetic term of the Lagrangian
and the scalar potential. We first assume that no direct coupling between $T$ and the matter fields exists,
except for the superpotential coupling to $\phi_1$ or $\vphi_1$ necessary to obtain the desired inflationary potential. 
The scalar kinetic term (\ref{skin_phi}) may then be expanded to first order in $\delta T$, to yield 
\beq
\label{skin}
\mathcal{L}_{B,{\rm kin}}=\,\frac{1}{\sqrt{3}}\delta T\phi_i\partial_{\mu}\partial^{\mu}\bar{\phi}^i + \frac{n_a}{\sqrt{3}}\delta T \vphi_a\partial_{\mu}\partial^{\mu}\bar{\vphi}^a  + {\rm h.c.} + \mathcal{O}\left(\Delta^2\right)  + \cdots
\eeq
%
The gauge-independent part of the scalar potential (\ref{spot_phi}) can also be expanded to reveal the interaction terms:
\beq\label{spot}
\begin{aligned}
\mathcal{L}_{B,{\rm pot}} &= - \frac{B^I_J}{\sqrt{3}} \delta T \Phi_I\bar{\Phi}^J - \frac{B^{IJ}_K}{2\sqrt{3}} \delta T \Phi_I\Phi_J\bar{\Phi}^K - \frac{B^I_{JK}}{2\sqrt{3}} \delta T \Phi_I\bar{\Phi}^J\bar{\Phi}^K -\frac{B^{IJK}_{L}}{6\sqrt{3}}\delta T \Phi_I\Phi_J\Phi_K\bar{\Phi}^L\\
&\quad\   -\frac{B^{I}_{JKL}}{6\sqrt{3}}\delta T \Phi_I\bar{\Phi}^J\bar{\Phi}^K\bar{\Phi}^L -\frac{B^{IJ}_{KL} + C^{IJ}_{KL}}{4\sqrt{3}}\delta T \Phi_I\Phi_J\bar{\Phi}^K\bar{\Phi}^L + {\rm h.c.} + \mathcal{O}(\Delta) + \cdots \, .
\end{aligned}
\eeq
Here the coefficients $B^{I_1I_2\dots}_{J_1J_2\dots}$ are as defined in (\ref{Bij}), 
and the $C^{IJ}_{KL}$ are sector-dependent functions of the bilinear coupling constants of the superpotential given by
\begin{align}
C^{IJ}_{KL} &=-\big(3+(n_I+n_J-3)(n_K+n_L-3)\big)W^{IJ}\bar{W}_{LK}\notag\\
&\quad\ +(n_I+n_M-3)\delta^I_LW^{JM}\bar{W}_{MK}. \label{Cijkl} 
\end{align}
Note that this expression assumes $n_i=1$ when $I$ represents an untwisted field (see below).
We have ignored the couplings to $\phi_1$ (or $\vphi_1$), due to the fact that this field, when coupled to $T$, 
possesses a mass equal to the inflaton mass $m$, and therefore the decay of $T$ to $\phi_1$ ($\vphi_1$) is kinematically forbidden. 

At tree level, the equation of motion for the conjugate matter fields may be substituted in (\ref{skin}). 
To quadratic order, the tree-level contribution from the \kahler potential for untwisted matter fields has the same form as that of twisted matter fields with unit modular weight, 
\beq
K\supset -3\log\left(T+\bar{T} - \sum_{i}\frac{|\phi_i|^2}{3}\right) = -3\log\left(T+\bar{T}\right) + \sum_{i}\frac{|\phi_i|^2}{T+\bar{T}} + \cdots
\eeq
Therefore, at this level we can define $n_i\equiv 1$, where $i$ runs over all untwisted matter fields. 
The effective Lagrangian, including the contributions from (\ref{skin}) and (\ref{spot}), can then be written as
\begin{align}\label{lseff}
\mathcal{L}_{B,{\rm eff}} &= -\frac{\delta T}{\sqrt{3}}(n_I + n_L - 3)W^{IL}\bar{W}_{LJ}\Phi_I\bar{\Phi}^J \notag\\
&\quad - \frac{\delta T}{2\sqrt{3}}(n_I+n_L -3)W^{IL}\bar{W}_{LJK}\Phi_I\bar{\Phi}^J\bar{\Phi}^K \notag \\
&\quad - \frac{\delta T}{2\sqrt{3}}(n_I+n_J+n_L -3)W^{IJL}\bar{W}_{LK}\Phi_I\Phi_J\bar{\Phi}^K \notag \\
&\quad - \frac{\delta T}{\sqrt{3}}(n_J+n_L -3)W^{JL}\bar{W}_{LK}\Phi_I\Phi_J\bar{\Phi}^I\bar{\Phi}^K \notag\\
&\quad - \frac{\delta T}{6\sqrt{3}}(n_I+n_L -3)W^{IL}\bar{W}_{LJKM}\Phi_I\bar{\Phi}^J\bar{\Phi}^K\bar{\Phi}^M\\
&\quad - \frac{\delta T}{4\sqrt{3}}(n_I+n_J+n_L -3)W^{IJL}\bar{W}_{LKM}\Phi_I\Phi_J\bar{\Phi}^K\bar{\Phi}^M \notag\\
&\quad - \frac{\delta T}{6\sqrt{3}}(n_I+n_J+n_K+n_L -3)W^{IJKL}\bar{W}_{LM}\Phi_I\Phi_J\Phi_K\bar{\Phi}^M \notag\\
&\quad - \frac{\delta T}{12\sqrt{3}}(n_I+n_J -3)\big(9+(n_I+n_J-1)(n_K+n_M-3)\big)W^{IJ}\bar{W}_{KM}\Phi_I\Phi_J\bar{\Phi}^K\bar{\Phi}^M \notag\\
&\quad + \cdots\notag
\end{align}
Under the assumption that the masses of all scalar matter fields are hierarchically smaller than the inflaton mass, 
$m_{I}\ll m$, the two-body decay rate can be computed immediately:
\beq
\Gamma(T\rightarrow \Phi_I\bar{\Phi}^J) = (n_I + n_L - 3)^2\frac{|W^{IL}\bar{W}_{LJ}|^2}{48\pi m M_P^2}\,,
\eeq
where a sum over the repeated index $L$ is implied. This rate is dependent on the matter sector to
which the decay products belong, and is weak-scale suppressed in the case of MSSM scalars.
For example, the rate for decay to two Higgs bosons is
\beq
\Gamma(T\rightarrow H_{u,d}\bar{H}^{u,d}) = (2n_H-3)^2\frac{|\mu_H|^4}{24\pi m M_P^2}\ ,
\eeq
where $\mu_{H}$ denotes the MSSM bilinear Higgs coupling. These two-body rates lead to an extremely low reheating temperature:
for an inflaton mass $m\sim 10^{-5}M_P$, and $\mu_H \sim 1$ TeV, $T_R\sim 10^{-1}$ eV.
In the three-body case, the decay to light scalars is given by the widths
\begin{align}
\Gamma(T\rightarrow \Phi_I\bar{\Phi}^J\bar{\Phi}^K) &= (n_I+n_L-3)^2\frac{|W^{IL}\bar{W}_{LJK}|^2m}{12(8\pi)^3M_P^2}\ ,\\
\Gamma(T\rightarrow \Phi_I\Phi_J\bar{\Phi}^K) &= (n_I+n_J+n_L-3)^2\frac{|W^{IJL}\bar{W}_{LK}|^2m}{12(8\pi)^3M_P^2}\ .
\end{align}
In particular, the decay to the neutral $d$-type Higgs and the left and right stops has the rate
\beq\label{Htt}
\Gamma(T\rightarrow \bar{H}_d^0\bar{\tilde{t}}_R \tilde{t}_L,\, H_d^0\tilde{t}_R \bar{\tilde{t}}_L) = \left((2n_H-3)^2+(2n_t+n_H-3)^2\right)\frac{|\mu_H y_t|^2m}{4(8\pi)^3M_P^2}\,,
\eeq
where $y_t$ denotes the top Yukawa coupling. The corresponding reheating temperature is also low, in the MeV range.
The rates corresponding to four-body decays are the largest, despite the phase-space suppression. The decay width
\beq
\Gamma(T\rightarrow \Phi_I\Phi_J\bar{\Phi}^K\bar{\Phi}^M) = (n_I+n_J+n_L-3)^2\frac{|W^{IJL}\bar{W}_{LKM}|^2m^3}{72(8\pi)^5M_P^2} \, ,
\eeq
for which we have disregarded the bilinear couplings, implies the following decay rate to four stops
\beq
\Gamma(T\rightarrow \tilde{t}_R\tilde{t}_L\bar{\tilde{t}}_R\bar{\tilde{t}}_L) = (2n_t+n_H-3)^2\frac{|y_t|^4m^3}{8(8\pi)^5M_P^2}\,
\label{4stops}
\eeq
which corresponds to
\beq
T_R = |2n_t+n_H-3|(10^{7}\,{\rm GeV})g^{-1/4}|y_t|^2(\frac{m}{10^{-5}M_P})^{3/2} \, .
\label{T4}
\eeq
Thus, as long as the matter fields do not reside in the untwisted sector (for which $n_i = 1$ and the rate 
vanishes), we obtain an adequate reheating temperature.  
The preceding rates may be modified if there
are direct couplings between the modulus $T$ and the matter sector. A multiplicative coupling of the form
\beq
W\supset g(T)W_{\rm M}(\phi,\vphi)
\eeq
respects the form of the inflaton potential in the absence of linear terms in $W_{M}$. 
Assuming for simplicity that $g(1/2)=1$, the addition of the factor $g(T)$ to the matter superpotential 
results in rates proportional to those obtained for constant $g(T)$, for a given sector. In particular, 
for the effective Lagrangian (\ref{lseff}) it amounts to the substitution
\beq
(n_{I_1}+\cdots+n_L-3)W^{I_1I_2\dots L}\bar{W}_{LJ_1J_2\dots}\longrightarrow \ \big(n_{I_1}+\cdots+n_L-3+ g'(1/2)\big)W^{I_1I_2\ldots L}\bar{W}_{LJ_1J_2\ldots}.
\eeq
Therefore, the decay rates shown previously are enhanced by a factor of $|g'(1/2)|^2$, at most.

\subsubsection{Decays to matter fermions}

The direct decays of the volume modulus $T$ to fermions are determined by the couplings 
arising from the fermion kinetic term, the fermion mass matrix and the fermion-scalar 
interactions of the supergravity Lagrangian. The relevant couplings stemming from the 
fermion kinetic term (\ref{fer_kin}) are shown in (\ref{T_to_f_kin}) below. Similarly to the scalar case, 
the interactions are diagonal with respect to matter field indices:
\beq\label{T_to_f_kin}
\mathcal{L}_{F,{\rm kin}} =  -\frac{i}{2\sqrt{3}}\delta T \, \bar{\chi}_{iL}\gamma^{\mu}\partial_{\mu}\chi^i_L -\frac{i\,n_a}{2\sqrt{3}} \delta T \, \bar{\chi}_{aL}\gamma^{\mu}\partial_{\mu}\chi^a_L + {\rm h.c.} + \mathcal{O}(\Delta^2) + \cdots \, .
\eeq
The interaction terms derived from the fermion mass matrix and the fermion-scalar interactions (\ref{fer_ints}) correspond to
\begin{align}
\mathcal{L}_{F,{\rm int}}
=&\ \frac{i}{2\sqrt{3}}\bar{\chi}_{iL}(\slashed{\partial}\delta T )\chi_{L}^i + \frac{i\,n_a}{2\sqrt{3}}\bar{\chi}_{aL}(\slashed{\partial}\delta T )\chi_{L}^a + \frac{\sqrt{3}}{2}\delta T \,W^{IJ} \bar{\chi}_{IL}\chi_{JR} \notag\\
&+ \frac{\sqrt{3}}{2}\delta T \,W^{IJK}\Phi_K\bar{\chi}_{IL}\chi_{JR} + {\rm h.c.} + \mathcal{O}(\Delta^2) + \cdots \, .
\label{Lfint}
\end{align}
In analogy with the scalar case, at tree level the equation of motion for the fermion fields may be substituted in 
(\ref{T_to_f_kin}) and (\ref{Lfint}). Additionally, one must consider the fermion-dependent part of the equation of motion for the scalar fields in (\ref{skin}). Identifying $n_i=1$ for all untwisted matter fields, 
the effective interaction Lagrangian for fermion decays can be written as
\beq
\mathcal{L}_{F,{\rm eff}} = -\frac{\delta T}{2\sqrt{3}}(n_I+n_J-3)W^{IJ}\bar{\chi}_{IL}\chi_{JR} - \frac{\delta T}{2\sqrt{3}}(n_I+n_J+n_K-3)W^{IJK}\bar{\chi}_{IL}\chi_{JR}\Phi_{K} + \cdots\, .
\eeq
Assuming negligible masses for all final states, $m_{I}\ll m$, the rates for two-body decays to matter fermions take the form
\beq
\Gamma(T\rightarrow\bar{\chi}_I\chi_J)= (n_I+n_J-3)^2\frac{|W^{IJ}|^2m}{192\pi M_P^2}.
\eeq
which are (1/4) times the rate for three-body decays into scalars. 
The dominant rates are for three-body decays involving two fermions and one matter scalar are,
\beq\label{Tpcbc}
\Gamma(T\rightarrow \bar{\chi}_I\chi_J\Phi_K) = (n_I+n_J+n_K-3)^2\frac{|W^{IJK}|^2m^3}{36(8\pi)^3M_P^2}.
\eeq
which are non-vanishing in the MSSM so long the fields are twisted with weights $n_i \ne 1$. In particular, in the case of the top quark it implies the decay rate
\beq\label{THtt_f}
\Gamma(T\rightarrow H_u^0t_L\bar{t}_R, \, \tilde{t}_L\tilde{H}_u^0\bar{t}_R,\,\bar{\tilde{t}}_R t_L\tilde{H}_u^0) = (2n_t+n_H-3)^2\frac{|y_t|^2m^3}{12(8\pi)^3M_P^2}.
\eeq
which is somewhat larger than the four-scalar decay rate (\ref{4stops}) because of the three-body phase-space factor.

\subsubsection{Decays to supersymmetry-breaking moduli and the gravitino}

The volume modulus $T$ can also decay into the moduli responsible for the breaking of supersymmetry,
with an amplitude that depends on the details of the inflationary and supersymmetry-breaking sectors.
We consider first breaking by a hidden-sector untwisted matter field $z$, with the cubic superpotential (\ref{w_susy_1}).
The direct decay is mediated by the effective Lagrangian
\beq\label{T_to_phi2}
\mathcal{L}_{z} = -\frac{\delta T}{\sqrt{3}} (4 + \gamma) \,m_{3/2}^2 z \bar{z} + {\rm h.c.}  + \mathcal{O}(\Delta^2) + \cdots \, ,
\eeq
where $\gamma$ is a constant that depends on the inflationary model:
$\gamma= 8$ for the $(T,\phi_1)$ superpotential (\ref{tph_w}), 
and $\gamma = 6$ for the $(T,\vphi_1)$ superpotential (\ref{tvph_w}). 
All terms shown explicitly in (\ref{T_to_phi2}) are comparable and lead to the decay rate
\beq
\Gamma(T\rightarrow z\bar{z}) = \frac{(4+\gamma)^2 \Delta^4 m^3}{48\pi M_P^2} \, .
\label{Tphi2phi2}
\eeq
The coupling of $T$ to $z$ in the effective potential implies that, 
at the global supersymmetry breaking minimum, $\langle G^T\rangle \neq 0$. 
Therefore, the direct decay of $T$ to the gravitino is allowed. In the unitary gauge, 
this process is in general mediated by the Lagrangian (\ref{gravitino_L}). 
In the present case of supersymmetry breaking by $z$, 
the couplings are suppressed by the ratio of the gravitino mass to the inflaton mass, $\Delta$:
\beq
\mathcal{L}_{3/2,z} = -\frac{\sqrt{3}}{16}(82-13\gamma)\Delta^2\left[\frac{1}{2}\,\epsilon^{\mu\nu\rho\sigma} \bar{\psi}_{\mu}\gamma_{\nu}\psi_{\rho}\partial_{\sigma}\delta T  -im_{3/2}\delta T \,\bar{\psi}_{\mu}\sigma^{\mu\nu}\psi_{\nu}\right]  + \mathcal{O}(\Delta^4)+ \cdots \, .
\eeq
The decay rate can be readily evaluated to yield
\beq
\Gamma(T\rightarrow \psi_{3/2}\psi_{3/2}) \simeq (82-13\gamma)^2\frac{ \Delta^2 m^3}{768\pi M_P^2} \, .
\label{Tpsipsi}
\eeq
The decay widths (\ref{Tphi2phi2}, \ref{Tpsipsi}) are suppressed by powers of the ratio of the 
gravitino mass to the inflaton mass, $\Delta$, relative to the three-body matter decays (\ref{THtt_f}),
\beq
\frac{\Gamma(T\rightarrow \phi_2\bar{\phi}^2)}{\Gamma(T\rightarrow H_u^0t_L\bar{t}_R)}\sim 10^{3}\Delta^4\ , \qquad 
\frac{\Gamma(T\rightarrow \psi_{3/2}\psi_{3/2})}{\Gamma(T\rightarrow H_u^0t_L\bar{t}_R)}\sim 10^{3}\Delta^2\,,
\eeq
and hence are relatively unimportant for reheating.

One can also consider the scenario in which supersymmetry is broken by a strongly-stabilized 
Polonyi modulus in the twisted sector with superpotential (\ref{mu_pol}). In this case, the couplings between the inflaton 
$T$ and the Polonyi field $z$ are given by
\beq
\mathcal{L}_{z}= -5\sqrt{3}\,m_{3/2}^2\delta T zz -4\sqrt{3}\,m_{3/2}^2\delta T \bar{z}\bar{z} - 12\sqrt{3}\,\frac{m_{3/2}^2}{\Lambda_z^2}\, \delta T z\bar{z} + {\rm h.c.}  + \mathcal{O}(\Delta^4) + \cdots \, .
\eeq
Since $\Lambda_z\ll 1$, the dominant decay channel corresponds to $z\bar{z}$, with a rate
\beq
\Gamma(T\rightarrow z\bar{z})=\frac{27}{\pi}\frac{\Delta^4 M_P^2 m^3}{\Lambda_z^4} \, .
\eeq
The decay of $T$ to the gravitino in the Polonyi scenario is mediated by the following couplings:
\beq
\mathcal{L}_{3/2,z} = -\frac{3\sqrt{3}}{16}(4-3\bar{\gamma})\Lambda_z^{2\bar{\gamma}}\Delta^2\left[\frac{1}{2}\,\epsilon^{\mu\nu\rho\sigma} \bar{\psi}_{\mu}\gamma_{\nu}\psi_{\rho}\partial_{\sigma}\delta T  + i m_{3/2} \delta T \,\bar{\psi}_{\mu}\sigma^{\mu\nu}\psi_{\nu}\right] + \cdots \, ,
\eeq
where now $\bar{\gamma}=0$ for the $(T,\vphi_1)$ superpotential (\ref{tvph_w}), 
and $\bar{\gamma}=1$ for the $(T,\phi_1)$ superpotential (\ref{tph_w}). 
In the latter case the amplitude is further suppressed by the factor $\Lambda_z^2$. 
The width is then given by
\beq
\Gamma(T\rightarrow \psi_{3/2}\psi_{3/2}) \simeq (4-3\bar{\gamma})^2\left(\frac{\Lambda_z}{M_P^2}\right)^{4\bar{\gamma}}\frac{3 \Delta^2m^3}{256\pi  M_P^2} \, .
\eeq
It is straightforward to verify that the decays of $T$ to the Polonyi modulus and to the gravitino in this scenario are
negligible relative to the matter decay (\ref{THtt_f}),
\beq
\frac{\Gamma(T\rightarrow z\bar{z})}{\Gamma(T\rightarrow H_u^0t_L\bar{t}_R)}\sim 10^{6}\left(\frac{\Delta}{\Lambda_z/M_P}\right)^4\ , \qquad 
\frac{\Gamma(T\rightarrow \psi_{3/2}\psi_{3/2})}{\Gamma(T\rightarrow H_u^0t_L\bar{t}_R)}\sim 10^{3} \Delta^2\left(\frac{\Lambda_z}{M_P^2}\right)^{4\bar{\gamma}}\,.
\label{zzgg}
\eeq

The decays to a single gravitino and a fermion belonging to a chiral multiplet are mediated by the 
interaction terms (\ref{sing_grav}). It is straightforward to show that the amplitudes for the decays 
with a final-state matter fermion vanish up to $\mathcal{O}(\Delta^2)$. 
The only non-vanishing couplings with $T$ are those with the inflatino and the $\phi_1$ or $\vphi_1$-ino. 
The corresponding amplitudes are dependent on the supersymmetry-breaking mechanism.
However, in all cases it can be shown that the decay rates to kinematically-allowed final-state
mass eigenstates are suppressed by a factor of $\Delta^2$: $\Gamma\sim \Delta^2(m^3/M_P^2)$, due to
the mass degeneracy $m-m_{\chi} \sim m_{3/2}$.

In the absence of a direct coupling of $T$ to the gauge degrees of freedom, i.e.,
$f_{\alpha\beta}^T=0$, where $f_{\alpha\beta}$ is the gauge kinetic function, 
the total decay rate of the inflaton is the sum of the rates previously shown. 
The largest width corresponds is that to two matter fermions plus a matter scalar, 
(\ref{THtt_f}), which implies the reheating temperature
\beq\label{TR_fer}
T_R = (10^8\ {\rm GeV})\,|y_t(2n_t+n_H-6)|\left(\frac{g}{915/4}\right)^{-1/4}\left(\frac{m}{10^{-5} M_P}\right)^{3/2}.
\eeq

\subsubsection{Decays to gauge bosons and gauginos}

The inflaton $T$ can decay to gauge fields and gauginos through a coupling in the 
gauge kinetic function $f_{\alpha\beta}(T)$, which, as was mentioned before, 
is a generic feature of heterotic string effective field theories~\cite{ADEH,gkf}. 
The supergravity Lagrangian terms containing the relevant interactions are given by (\ref{gaugeL}), 
disregarding contributions suppressed by the gaugino masses. 
The decay width to the canonically-normalized gauge boson pairs is readily evaluated, resulting in
\beq
\Gamma(T\rightarrow gg) = \frac{d_{g,T}^2}{32\pi}\left(\frac{N_G}{12}\right)\frac{m^3}{M_P^2}\,,
\eeq
where $N_G$ is the number of final states: $N_G=12$ for the Standard Model, 
and $d_{g,T}$ has been defined in (\ref{dgT}). The corresponding reheating temperature is
\beq
T_R =  (3\times10^{9}\ {\rm GeV})\, d_{g,T}\left(\frac{N_G}{12}\right)^{1/2} \left(\frac{g}{915/4}\right)^{-1/4}\left(\frac{m}{10^{-5} M_P}\right)^{3/2}.
\label{Tgg}
\eeq
The coefficient $d_{g,T}$ might well be $\mathcal{O}(1)$, 
e.g., for a gauge kinetic function linear in $T$ with $\mathcal{O}(1)$ coefficients,
in which case all other decay channels of the volume modulus $T$ would be overwhelmed by the decays to gauge bosons,
and the reheating temperature would be large. The effective reheating temperature generated by
decays into gauge bosons would exceed that due to decays into matter particles, (\ref{TR_fer}), for any $d_{g,T} \gtrsim {\cal O}(1/30)$.

On the other hand, the decays of $T$ to gauginos are subdominant.
Our results differ from the treatment of \cite{Nakamura:2006uc},
in that in our case the mass of the modulus $T$ is determined not by the bilinear coupling $W^{TT}$, 
which has a vanishing vev, but by the coupling $W^{T\phi_1}$ or $W^{T\vphi_1}$. 
This results in an amplitude for decay to gauginos that is suppressed by $\Delta$ relative to the 
amplitude for the decay to gauge bosons. Assuming for simplicity that $f(T)$ is a holomorphic function 
with real coefficients, the corresponding decay rate is
\beq
\Gamma(T\rightarrow \tilde{g}\tilde{g}) = \frac{d_g^2 \Delta^2}{16\pi}\left(\frac{N_G}{12}\right)\frac{m^3}{M_P^2}\,.
\eeq
A similar suppression for the decay to gauginos was seen in \cite{klor}.

\section{Summary and Prospects}

We have considered in this paper various aspects of no-scale inflation,
considering two main classes of models: those in which the inflaton is
identified with an untwisted matter field $\phi$, and those in which the inflaton
is identified with the compactification volume modulus $T$. We have focused on
two important phenomenological issues: possible patterns of soft
supersymmetry breaking, and inflaton decays and the related
reheating temperature of the Universe subsequent to inflation.
We have considered in Section~4 various possible mechanisms for supersymmetry
breaking, including via the volume moduli and the Polonyi mechanism.
These mechanisms yield many possibilities for the soft
supersymmetry-breaking parameters effective low-energy theory.
In general, the patterns of soft supersymmetry breaking for the
untwisted and twisted matter sectors are different. For example,
no-scale, CMSSM or mSUGRA boundary conditions are natural possibilities in the
untwisted sector, whereas in the twisted sector the soft supersymmetry-breaking
parameters are not universal in general, since they depend on the modular weights
of the fields. As usual, the gaugino masses would in general arise from
a non-minimal gauge kinetic function or through loop effects via anomalies.
Observation of supersymmetric particles at the LHC or elsewhere followed
by studies of the pattern of supersymmetry breaking could give valuable
insights into the form of no-scale inflationary model and the assignments
of matter particles as well as the inflaton.

In Section~5 we have considered inflaton decays in the same two classes of models,
namely when the inflaton is the untwisted matter field $\phi$, and when the inflaton is
the volume modulus field $T$.
The reheating temperature could in principle be larger in the $\phi$ inflaton case, namely ${\cal O}(10^{15})$~GeV,
if there is an ${\cal O}(1)$ trilinear superpotential coupling between $\phi$ and light matter fields,
as might occur in a neutrino inflation
scenario, see (\ref{phisneu}) and (\ref{phineu}). A similar reheating temperature could in principle also be generated by decays into gravitino
and modulino pairs, see (\ref{phigneu}). However, the gravitino problem imposes a non-trivial upper limit
on the reheating temperature, and hence on the possible trilinear superpotential,
moduli and gravitino couplings. On the other hand, the reheating temperature
is naturally considerably smaller in the $T$ inflaton case, namely ${\cal O}(3 \times 10^8)$~GeV,
with the dominant decays being into three-body matter final states, see (\ref{THtt_f}), whereas decays into
gravitinos are expected to be much smaller (\ref{zzgg}). We note, however,
that in both scenarios decays into gauge bosons may also yield
reheating temperatures as large as $10^{10}$~GeV, depending on the form of the gauge kinetic function,
see Eqs.~(\ref{phigg}) and (\ref{Tgg}). This provides a possible link between the supersymmetry-breaking
mechanism for generating $m_{1/2} \ne 0$ and the thermal history of the Universe.

As commented in the Introduction, the values of the cosmic microwave background observables $n_s$ and $r$ are sensitive
to the number of e-folds during inflation, $N^*$, which is in turn sensitive to the reheating temperature.
Thus, there is in principle a connection between accelerator physics and inflationary cosmology
via supersymmetry breaking, which may also cast light on the nature of string compactification.
However, detailed exploration of this fascinating connection lies beyond the scope of this paper.

\section*{Acknowledgements}

The work of J.E. was supported in part by the London Centre for Terauniverse Studies
(LCTS), using funding from the European Research Council via the Advanced Investigator
Grant 267352 and from the UK STFC via the research grant ST/J002798/1.
The work of D.V.N. was supported in part by the DOE grant DE-FG02-13ER42020. 
The work of M.A.G.G. and
K.A.O. was supported in part by DOE grant DE-SC0011842  at the University of
Minnesota.

\end{document}